\documentclass[preprint,amsmath,amssymb,aps,preprintnumbers,superscriptaddress,physrev,11pt]{revtex4-2}
\usepackage[utf8]{inputenc}
\usepackage{amsmath}
\usepackage{amssymb}
\usepackage{graphicx}
\usepackage[margin=1in]{geometry}

\begin{document}

\title{A Gap No More: Mechanism for Non-Nuclear Energy to Fill in the Black Hole Mass Gap}
\author{Joshua Ziegler}
\email{jjziegler@utexas.edu}
\affiliation{Texas Center for Cosmology and Astroparticle Physics, Weinberg Institute for Theoretical Physics,
Department of Physics, University of Texas, Austin, Texas 78751, USA}
\author{Katherine Freese}
\email{ktfreese@utexas.edu}
\affiliation{Texas Center for Cosmology and Astroparticle Physics, Weinberg Institute for Theoretical Physics,
Department of Physics, University of Texas, Austin, Texas 78751, USA}
\affiliation{The Oskar Klein Centre, Department of Physics, Stockholm University, AlbaNova, SE-10691
Stockholm, Sweden}
\affiliation{Nordic Institute for Theoretical Physics (NORDITA), 106 91 Stockholm, Sweden}

\begin{abstract}

Standard stellar evolution models predict that black holes in the range of approximately $50 - 140 M_\odot$ should not exist directly from stellar evolution. This gap appears because stars with masses between 100 and 240 $M_\odot$ are expected to undergo a pair instability supernova and leave behind no remnant, or a pulsational pair instability supernova and leave behind a remnant much smaller than their initial stellar mass. However, black holes have been discovered by the LIGO/Virgo collaboration within this mass range. In previous work\cite{ziegler2020}, we used the stellar evolution code MESA to show that the addition of non-nuclear energy (such as from annihilation of dark matter) could alter the evolution of a 180 $M_\odot$ star so that the observed black holes could be produced from isolated stars. In this paper, we extend this analysis to stars of other masses, and find that sufficient amounts of non-nuclear energy can allow any star to avoid pair instability, and could produce a black hole of mass comparable to the initial stellar mass. In addition, we produce examples of the type of black hole initial mass function that can be produced from this mechanism. These illustrative examples suggest that adding non-nuclear energy to stars offers a way to fully close the mass gap. 

\end{abstract}

\preprint{UTWI-25-2022, NORDITA-2022-101}
\maketitle

\section{Introduction}

Since the first direct observation of gravitational waves in 2015 \cite{ligo2015} by the LIGO collaboration, the number of identified black holes has skyrocketed. In the coming years, that catalogue is expected to grow even more. While individual observations can tell us a great deal about gravity or the physics of merging black holes, the statistics of the black holes that LIGO, Virgo, KAGRA, etc, will be able to observe may offer significant insights toward other outstanding questions in astrophysics. For example, in this paper, we explore how black hole statistics may provide a way to probe stellar processes, and potentially the nature of dark matter.

The identification and characterization of the properties of dark matter is among the most enduring problems faced in cosmology. First proposed almost a century ago, the idea of a non-interacting mass component to the universe is not new. However, despite extensive effort to determine the properties of dark matter, it remains largely a mystery. Despite making up a quarter of the energy of the universe, the low background density and cross-section of dark matter makes it very difficult to study. As a result, the hunt for dark matter has begun to involve ever more inventive search techniques, powerful detectors, and creative ways to use existing data.

For example, the data being collected by gravitational observatories like the Laser Interferometer Gravitation Observatory (LIGO) and its partners Virgo and KAGRA, offer a new window with which to potentially probe properties of dark matter (see, for example\cite{Bertone_2020}). Many of the techniques proposed to study new physics with gravitational waves use the gravitational waveforms themselves\cite{Chia2022,Coogan_2022}.
However, gravitational wave observatories offer another way to probe new physics: the properties of the compact objects which source the gravitational waves.

As of the recently completed O3 run, the LIGO/Virgo/KAGRA collaborations have identified 90 events which are consistent with the mergers of compact objects like black holes and neutron stars\cite{2111.03606}. Upcoming upgrades to the detectors and analysis pipeline promise to increase that number several-fold over the coming years. As the number of black holes detected by these observatories grows, their properties can be used to construct the black hole initial mass function (BHIMF). The BHIMF describes the number of black holes formed as a function of the mass of those black holes, i.e. $N(M)$, and can in principle be predicted from stellar evolution studies (e.g. \cite{Sicilia_2022}). As a result, gravitational wave observatories can, through the BHIMF, provide some insights into stellar evolution. However, in general, the insights that can be gained about stellar evolution in this way are limited by the complexity inherent in stellar evolution. We are ultimately interested in probing new physics with the BHIMF, and the effects of the new physics will generally be quite difficult to distinguish from the effects of various stellar effects, some of which are poorly understood.

However, there is at least one region of the black hole mass function where the backgrounds from standard stellar evolution are small enough that this degeneracy problem may be manageable: the (upper) black hole mass gap. Standard models of stellar evolution posit that stars in the range of masses $150 M_\odot$ to $260 M_\odot$ undergo pair instability supernovae (PISNe)\cite{woosley2002, woosley2014, woosley2017, woosley2019, belczynski2016, spera2017}. This type of supernova has the characteristic feature of completely destroying the progenitor, and leaving behind no remnant. As a result, there are predicted to be no black holes formed directly from stellar evolution in the range of masses $50 M_\odot$ to $140 M_\odot$, which forms the black hole mass gap. This phenomenon is particularly robust in standard stellar evolution, and while the boundaries of the gap are uncertain and can vary between different authors by up to $10 M_\odot$, a gap is strongly expected to exist.

However, in 2019, the LIGO and Virgo collaborations discovered a merger of two black holes with best-fit masses $66~M_\odot$ and $85~M_\odot$\cite{ligo2020, ligo2020a}. These masses would place both black holes squarely inside the mass gap. 
Since the discovery of GW190521, several explanations for its existence have been proposed. Some authors have proposed alternative parameter estimation from the gravitational wave data \cite{fishbach2020, gayathri2020}, which can alter the inferred masses of the black holes involved. Others argue that the mass gap boundary can be shifted so that black holes with the observed masses are not ruled out \cite{branch, farrell2020}. Still others accept that the black holes exist in the mass gap, and try to explain how they could be seen there. For instance, one proposed solution is that the observed merger involved second-generation black holes \cite{fragione2020} or black holes formed after two stellar progenitors collided\cite{carlo2019,carlo2019short,vanSon2020}. Primordial black holes could also explain the observed black holes, either directly \cite{carr2019} or through mergers \cite{luca2020}. And extensions of the standard model offer a range of potential explanations\cite{sakstein2020}, from axions\cite{croon2020} to \cite{straight2020}, and more. Clearly there are many ways for black holes to appear in the mass gap, but these mechanisms will generally predict different overall behaviors in the BHIMF.

To that end, we explore here how a particular model which can explain black holes in the mass gap may show up in the BHIMF. In recent work\cite{ziegler2020}, we were able to show that stars could avoid evolving to a PISN even if they had an initial mass between $150-260~M\odot$ if there was non-nuclear energy (NNE) added to the star. In particular, we showed that a star with mass $180 M_\odot$, where approximately 60\% of the energy of the star comes from NNE, would evolve to a (core-collapse) supernova progenitor with mass at least $119 M_\odot$. We now extend this analysis to a wider range of stellar masses and NNE rates, and are therefore able to explore the signatures which would appear in the BHIMF from this class of models. Furthermore, by extending our analysis to a wider range of stellar masses, we can make general statements about how the inclusion of NNE can change stellar evolution, in invisible and potentially visible ways.

The remainder of this paper is laid out as follows: in section \ref{evolution}, we provide an overview of the standard evolution of stars with initial mass on the order of $100 M_\odot$. In section \ref{modeling}, we discuss the simulations we performed to model the evolution of such stars. In section \ref{results}, we compare the evolution of different stellar masses with different amounts of NNE, and provide general characteristics. In section \ref{dark matter}, we provide a model for the source of NNE in the form of dark matter, as well as some proposals for how the energy production could conceivably be realized. In section \ref{BHIMF}, we propose possible examples of the BHIMF we could see under this model. Finally, we conclude in section \ref{conclusion}.

\section{Standard Model of Stellar Evolution}
\label{evolution}

Our goal is to understand how new physics could affect the evolution of stars, so we describe here the evolution of massive stars based on known physics to use as a comparison for our results below. Throughout this description, we focus specifically on the evolution of stars with masses between $70$ and $300 \, M_\odot$. In order for stars to maintain this mass throughout their lifetimes, they must experience weak stellar winds, which typically implies that the metallicity in the stars is quite low. What we describe here is based on modeling of stars with metallicity $Z=0$, or primordial chemical composition, similar to that done in \cite{woosley2002}. 

All stars begin their evolution as clouds of gas which contract under the force of gravity. This contraction causes the temperature and density within the gas to increase, until eventually protons can overcome Coulomb repulsion in order to begin hydrogen fusion. The high mass stars we consider typically will continue contracting until helium is also fusing into carbon in the core in order to produce enough energy to balance against gravity. Over a period of a few million years, the star gradually depletes the hydrogen in an inner convective region, while the outer layers of the star retain their hydrogen content. (While helium is also depleted, the hydrogen fuses more rapidly, and so even once all of the hydrogen is gone, helium still remains.) At this point, the difference in chemical composition between the envelope and the core establishes a barrier to mixing that is possible to overcome with standard stellar processes only in specific situations. Because of the lack of convective mixing between these two regions, we can use this boundary to establish a well-defined envelope (hydrogen dense material outward of the convective boundary), and core (hydrogen-poor material inward of the boundary).

Following hydrogen depletion in the core, the star contracts and temperatures increase, until the energy from helium fusion is sufficient to support the star. For hundreds of thousands of years, as helium fusion progresses in the core, the star is largely unchanging, but gradually builds up an inner carbon-oxygen core. As helium is depleted in the core, the core contracts as before, but now the contraction leads the inner edge of the envelope to heat up as well. This heating can trigger hydrogen fusion at the surface of the helium shell. Likewise, helium fusion may occur at the boundary between the helium shell and the carbon-oxygen core. In some stars, this shell-burning can be associated with the process-known as dredge-up, which can allow for some mixing between the helium shell and the envelope. As a result, elements created within the core of the star, like carbon, nitrogen, and oxygen, can be dredged up into the envelope and eventually reach the surface, where they may be observable in stellar spectra. This dredge-up phenomenon, however, does not happen in all stars, but seems to happen only in stars at the low end of the mass range we consider.

In any case, as the temperature of the core increases, two processes become important. In terms of nuclear fusion processes, stars begin to undergo carbon and oxygen fusion, building up a range of elements from magnesium to silicon. At the same time, however, the temperature reaches a point where the equilibrium production of electrons and positrons from photons is no longer negligible. More specifically, the equilibrium of electrons, positrons, and photons shifts to favor a higher fraction of energy in electrons. Because photons produce more pressure than electrons for the same amount of energy, this shift lowers the pressure support within a star. While all stars with mass $M\gtrsim 8\, M_\odot$ reach these temperatures, lower mass stars tend to have a higher central density. The higher density (of electrons in particular) implies that quantum effects are not negligible and the production of electron-positron pairs is limited by phase space constraints. One way to quantify all of these effects is through the thermal adiabatic constant $\gamma = \partial \log P/ \partial\log \rho$.

By using the averaged values of $\gamma$ reached during a star's lifetime, we can distinguish four types of evolution, each of which occurs in a typical mass range.
\begin{enumerate}
   \item For stars with masses $M\lesssim 100 M_\odot$, the central density of the star is high enough, and temperature low enough, that $\gamma >4/3$
   throughout the stellar core while the star is primarily powered through carbon- and oxygen-burning. The star continues to fuse carbon and oxygen into heavier elements until this fusion is not sufficient to fully support the star. The star contracts and heats up, allowing nuclear reactions involving heavier nuclei to occur. This process continues until an iron core forms. As the core further increases in temperature, this iron can begin to fuse, but these iron reactions require energy from the star. As a result, the pressure within the star decreases, leading to a runaway collapse of the star. This collapse is halted by neutron degeneracy pressure, and the subsequent bounce leads to a type of supernova known as a core-collapse supernova. Depending on the exact properties of the supernova explosion, core collapse supernovae (of the massive stars we consider here) leave behind black holes whose masses are around half that of their progenitor.
   \item For stars with masses $\sim 100 M_\odot - 140 M_\odot$, higher temperature and lower density allows for a portion of the star to experience $\gamma<4/3$. This portion of the star is in a spherical annulus slightly off-center, rather than directly at the center. This portion of the star collapses inward, causing some of the fuel-rich outer layers of the core to have a high enough temperature to begin fusing. In particular, the fusion of oxygen and, to a lesser extent, silicon produce a powerful burst of energy. This energy burst is strong enough to unbind the outer layers of the star, and to cause a reduction of the density of the inner core. Gradually, the core re-contracts until it again begins fusing carbon and oxygen. Now a less massive object, what remains of the star has a lower temperature and higher density. This star may therefore be able to avoid producing electron-positron pairs in abundance, and consequently may evolve to a core collapse supernova. Alternatively, if the star is still massive enough to produce sufficient amounts of electron-positron pairs, it may undergo another explosive burst of oxygen fusion. This process can occur many times until eventually enough mass is lost that the star can avoid producing electron-positron pairs in abundance, leading to a core collapse supernova. This entire process is called a pulsational pair instability supernova (PPISN).
   \item For stars with masses $\sim 140 M_\odot - 250 M_\odot$, the temperature and density are such that at some point the entire core experiences $\gamma<4/3$. Consequently, when the core contracts as a result of the electron-positron pair production, there is a substantial burst of oxygen and silicon fusion. This explosion is powerful enough to completely unbind the star and cause it to explode without leaving behind any remnant. This explosion is referred to as a pair instability supernova (PISN).
   \item For more massive stars with masses $M \gtrsim 250 M_\odot$, the production of electron-positron pairs also occurs, causing a collapse of the star. However, the energy released by the fusion of oxygen and silicon is not sufficient to cause an explosion of the star. Instead, the star continues collapsing, and leads to a core-collapse supernova. Therefore, the mass of the black hole formed from progenitors in this mass range is approximately half the progenitor's mass.
\end{enumerate}

As a result of these different possible evolutionary paths, stars with masses between approximately $140-250~M_\odot$ do not leave behind any remnant. Furthermore, those with masses between $100-140~M_\odot$ leave behind a black hole with a mass significantly less than than the initial stellar mass. As a result there is a gap in the masses of black holes which can form directly from stellar evolution.

\section{Numerical Recipe}
\label{modeling}

To arrive at the above description of stellar evolution, we must make several assumptions about stellar structure. In general, the details of a star's internal structure will be inherently three-dimensional and will depend sensitively on stellar properties like mass, chemical composition, rotation rate, and magnetic fields. This problem is therefore quite challenging. However, we can simplify this in a few ways. First, because we are looking at very massive stars ($70 - 300 M_\odot$), we can assume the initial chemical composition is approximately primordial. Equivalently, we can assume that the initial metallicity is $Z \approx 0$. Higher initial metallicity implies a higher opacity, which powers a stronger stellar wind. As a result, stars with a higher metallicity tend to experience higher mass loss, and are unlikely to maintain the desired mass throughout their lifetimes.\cite{woosley2002}. We will also be assuming that the stars we consider are non-rotating. This assumption implies that magnetic fields are small, and we will assume that they are negligible. Furthermore, we assume that the stars we consider are isolated from any other systems. 

By assuming that stars are non-rotating and non-magnetic, we simplify the physical problem of their internal structure to be manifestly spherically symmetric, and described in detail in \cite{kippenhahn}. The only forces acting on the stellar material are gravitational and pressure. Gravitational forces attract stellar material radially inward, while gas pressure and radiation pressure provide an outward radial force. The balance of these two forces determines the internal structure of a star, and specifically the temperature, density, and pressure profiles, $T(r), \rho(r),$ and$ P(r)$, respectively, where $r$ is the distance from the center of the star. For most of a star's lifetime, the pressure which supports a star against gravitational contraction comes from nuclear fusion, which produces a luminosity profile $L(r).$ At any given time, we can relate these four quantities through the following system of equations, where for ease of computation, we have switched dependent variables to $M$, the stellar mass interior to a radius $r$\cite{kippenhahn}
\begin{align}
    \frac{\partial r}{\partial M} &= \frac{1}{4\pi r^2 \rho}\\
    \frac{\partial P}{\partial M} &= -\frac{GM}{4\pi r^2}\\
    \frac{\partial L}{\partial M} &= \epsilon\\
    \frac{\partial T}{\partial M} &= -\frac{GMT}{4\pi r^4 P}\nabla.
\end{align}
In this set of equations, besides the four functions which describe structural properties of the star, there are three additional functions, which are assumed to be constant for all stars. These are the functions $\rho(P, T, X)$, $\epsilon(T,\rho, X)$, and $\nabla(T,\rho,X)$. In each of these cases, $X$ describes the chemical composition at a particular location in the star, which for our purposes begins at primordial composition and evolves by nucleosynthesis. The equation of state $\rho(P, T, X)$ relates density at a given location in the star with the temperature and chemical composition there, but is fundamentally a property of the stellar material. The energy production rate $\epsilon(T,\rho, X)$ describes the amount of energy produced through nuclear reactions in a region with temperature $T$, density $\rho$, and chemical composition $X$. Finally, the function $\nabla(T,\rho,X) \equiv \frac{d \log T}{d\log P}$ includes a range of properties related to movement of energy through a star. For example, it can describe radiative transport of energy through opacity as well as convection. If we treat these three functions as well-established, the above system of equations can be solved to give the structure within a star, as defined by the functions $T(M), r(M), P(M),$ and $ L(M)$. 

Most of the evolution of the star can be determined by solving the above system of equations at sequential timesteps. Over any given time interval, the chemical composition throughout the star is propagated forward in time from an initial stellar structure based on nuclear reaction rates and convective processes. The stellar structure is then calculated for the resulting chemical composition using the above equations. This approach is sufficient to calculate the stellar evolution during periods of quasi-equilibrium, where changes to the structure are gradual and the time intervals are short compared to the timescale of any physical change to the star.

However, this equilibrium condition is not always satisfied. In particular, when gravitational forces dominate over outward pressure forces, the star can undergo dynamical contraction or collapse. In those scenarios, the above functions are insufficient. This can be resolved by amending the previous system of equations to 
\begin{align}
    \frac{\partial r}{\partial M} &= \frac{1}{4\pi r^2 \rho}\\
    \frac{\partial P}{\partial M} &= -\frac{GM}{4\pi r^2} - \frac{\partial^2 r}{\partial t^2}\frac{1}{4\pi r^2}\\
    \frac{\partial L}{\partial M} &= \epsilon - C_P \frac{\partial T}{\partial t} + \frac{\delta}{\rho}\frac{\partial P}{\partial t}\\
        \label{eqn:energyrate}
    \frac{\partial T}{\partial M} &= -\frac{GMT}{4\pi r^4 P}\nabla.
\end{align}
Here, the additional term in the second equation describes the change in gravitational energy caused by the change in structure of the star, which must now be balanced by pressure from the gas and radiation in the star. The two terms added in the third equation capture the energy released by changing the temperature and pressure of a gas, by introducing two functions $C_P$ and $\delta$ which describe thermodynamic properties of the stellar material. Combining these explicitly time-dependent factors and the gradual changes to chemical composition induced by nucleosynthesis can describe most of the evolution within a star, at least up to a supernova.

Throughout this paper, we explore stellar evolution using the MESA (Modules for Experiments in Stellar Astrophysics) code, version 12778\cite{Paxton2011, Paxton2013, Paxton2015, Paxton2018, Paxton2019}. This code determines the functions $P(M, t), r(M, t), L(M, t)$, and $T(M, t)$, as well as a customizable list of physical parameters on an adaptive grid. That is, the stellar interior is divided into a series of concentric spherical shells, whose thickness is variable, with thin shells near the center of the star and thicker shells in the envelope. In evaluating each stellar structure, the boundary conditions that $r(0) = 0$ and $L(0) =0$ are maintained at the center of the star, and $P(M_{surf})=0$ and $L_{surf} = 4\pi \sigma {r_{surf}}^2 {T_{surf}}^4$, where $\sigma$ is the Stefan-Boltzmann constant, are maintained at the surface.

The initial model is constructed from the known structure of a $30 M_\odot$ star with primordial composition. Mass is then added to the star in increments, and the structure recalculated, until the desired initial mass is reached. The composition is then adjust to realign with the desired composition, and the stellar structure recalculated. From this process, an initial structure of the star, evaluated at age 0, can be determined. At each subsequent timestep, changes to input parameters like the chemical composition distribution are applied to the previous model and a new stellar structure is calculated. If the structure cannot be calculated, the interval between timesteps is reduced, and the calculation attempted again. This process continues until one of three stopping criteria is met. Some models ended when successive failures to calculate the next timestep causes the time interval to reduce below $10^{-20}\, \mathrm{s}$. This can usually be resolved by adjusting numerical parameters used by MESA, including for example, the conditions which force the time intervals to change. Second, if the infall velocity in the center of the star exceeds $8\times10^8\, \mathrm{cm \, s^{-1}}$, we identify this as a core collapse supernova. And third, if the entire star exceeded the local escape velocity, the star is assumed to have undergone pair instability supernova.

In using MESA to model the stellar evolution, a range of tunable options are provided. We use a fixed initially primordial composition, and vary stellar masses over a range from $70 M_\odot$ to $300 M_\odot$. Specifically, we model stars with initial masses  $70, 110, 140, 180, 240$, and $300~M_\odot$. For the nuclear reaction network, we utilize the \texttt{approx-21} network in MESA, which includes 21 nuclear species and 116 reactions between them based on reaction rates from NACRE \cite{Angulo1999}, JINA REACLIB \cite{Cyburt2010}, plus additional tabulated weak reaction rates \cite{Fuller1985, Oda1994, Langanke2000}. We also include nuclear screening as prescribed by \cite{Chugunov2007} and include thermal neutrino loss as prescribed by \cite{Itoh1996}. This approach is designed to approximate the full nuclear network, and ensure that the overall energy output of nuclear reactions is consistent with true values. Equations of state are derived from OPAL \cite{Rogers2002}, SCVH \cite{Saumon1995}, PTEH \cite{Pols1995}, HELM \cite{Timmes2000}, and PC \cite{Potekhin2010} equations of state. To model the opacity, we use an opacity table generated by combining OPAL data \cite{Iglesias1993, Iglesias1996} with low-temperature data from \cite{Ferguson2005} and high-temperature data in the Compton-scattering dominated regime by \cite{Buchler1976}, as well as electron conduction opacities from \cite{Cassisi2007}. In addition, we include a stellar wind based on Brott \cite{brott2011}. To describe convection within the star, we use the Ledoux instability criterion \cite{ledoux1947} and an extension of mixing length theory (MLT+) with a mixing length parameter $\alpha = 2.0$, and include a small amount of exponential overshooting both above and below any convecting regions.

In the remainder of this paper, we explore the impact of adding NNE to the star. We do this by making use of the in-built \texttt{inject\_uniform\_extra\_heat} hook in MESA. Effectively, this adds an $\epsilon_{NNE}(M) = \mathrm{const.}$ which is constant as a function of the mass interior to a radius. We perform multiple runs for each mass of star, each including a different amount of NNE. The amounts of NNE we add varies between 0, which reproduces standard stellar evolution and can be used as a check on our approach, and a maximum value, above which the evolution of the star immediately leads to its dissolution. 

\section{Stellar Evolution with the Inclusion of Non-Nuclear Energy}
\label{results}

The addition of NNE to a star can cause a range of differences, particularly if the amount this NNE is comparable to the amount of energy from nuclear reactions. The main effects of adding NNE to a star can be understood simply in the following way:  adding NNE to a star causes less of its mass to be in the core and more in the envelope. These two changes in the mass distribution, both to the core and the envelope, contribute to further changes in the stellar structure and evolution. We discuss some of the consequences of the more massive
envelope in the next subsection. Meanwhile, the evolution of stars is primarily determined by the mass of their core, so stars with NNE added tend to evolve similarly to nuclear-only stars with a lower mass. 

\begin{figure}
    \centering
    \includegraphics[width=0.49\linewidth]{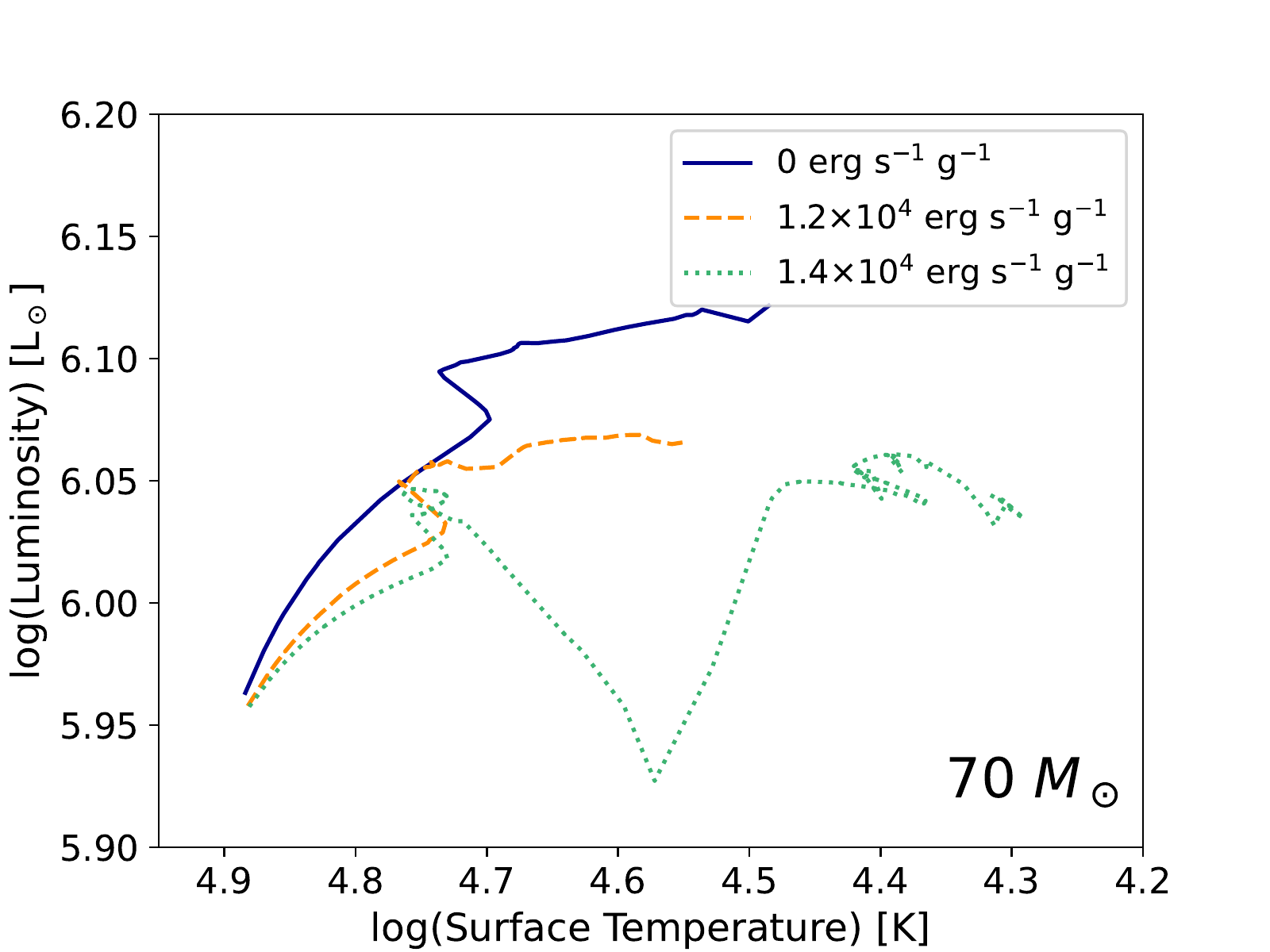}
    \includegraphics[width=0.49\linewidth]{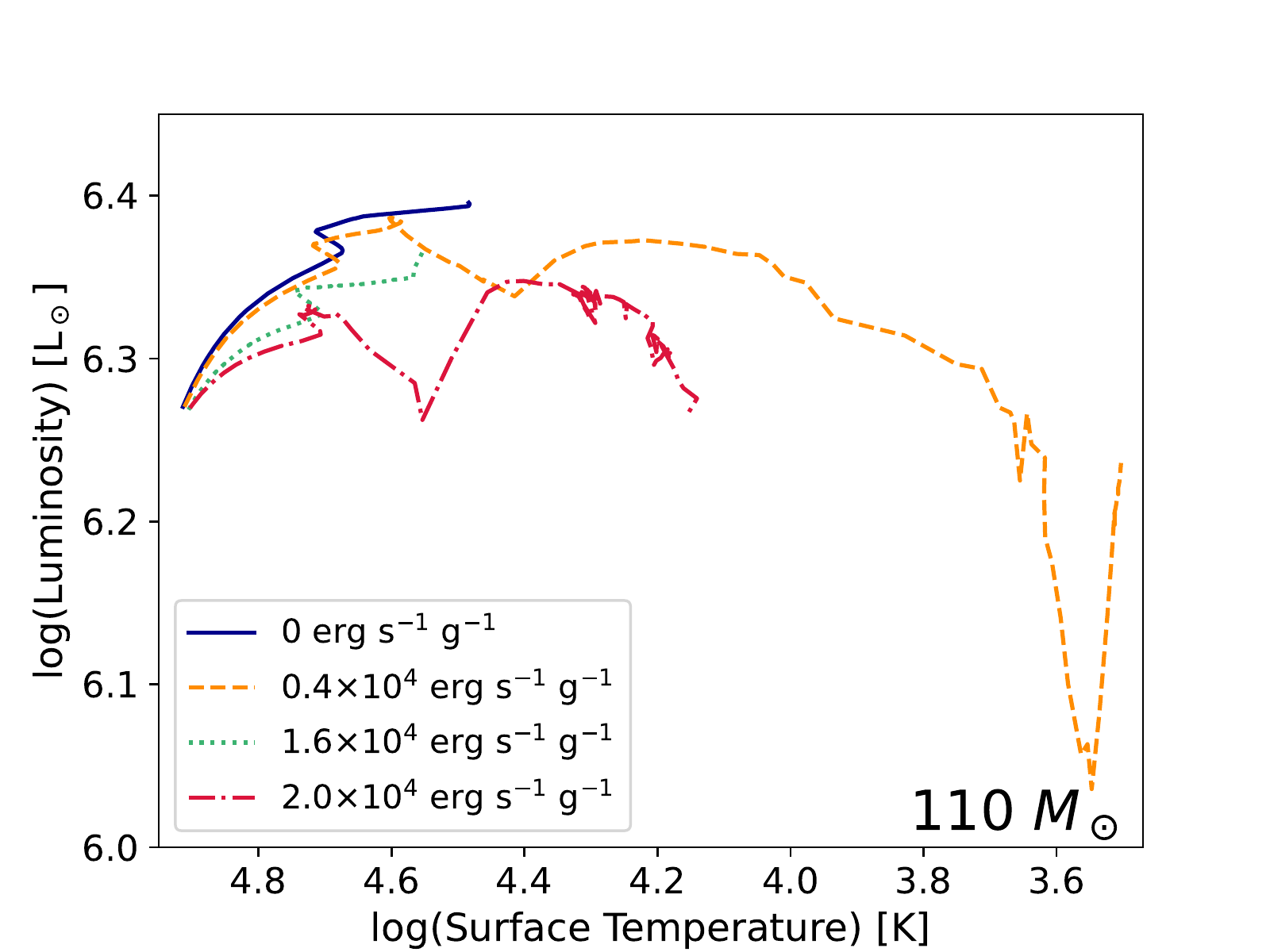}
    \includegraphics[width=0.49\linewidth]{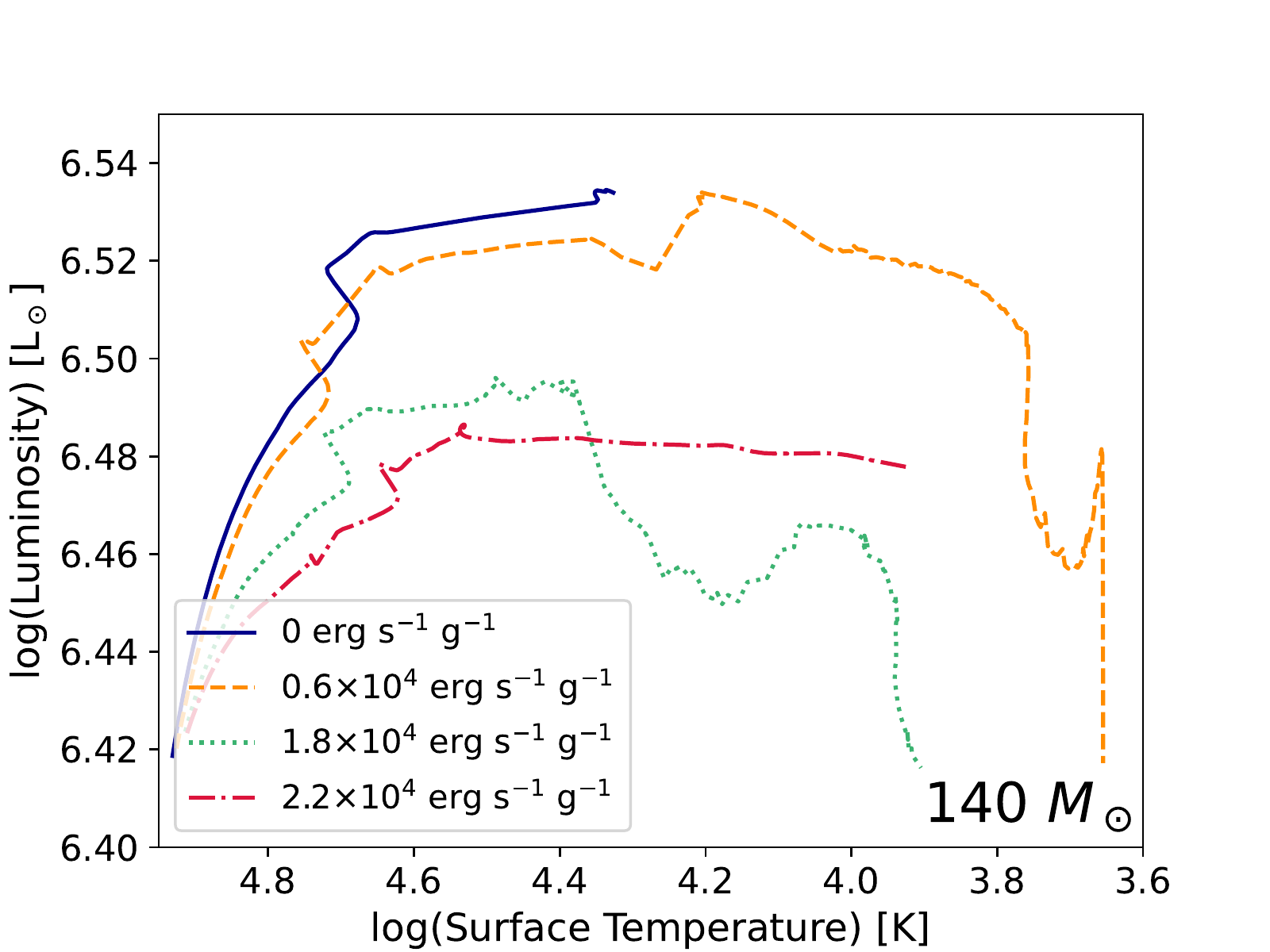}
    \includegraphics[width=0.49\linewidth]{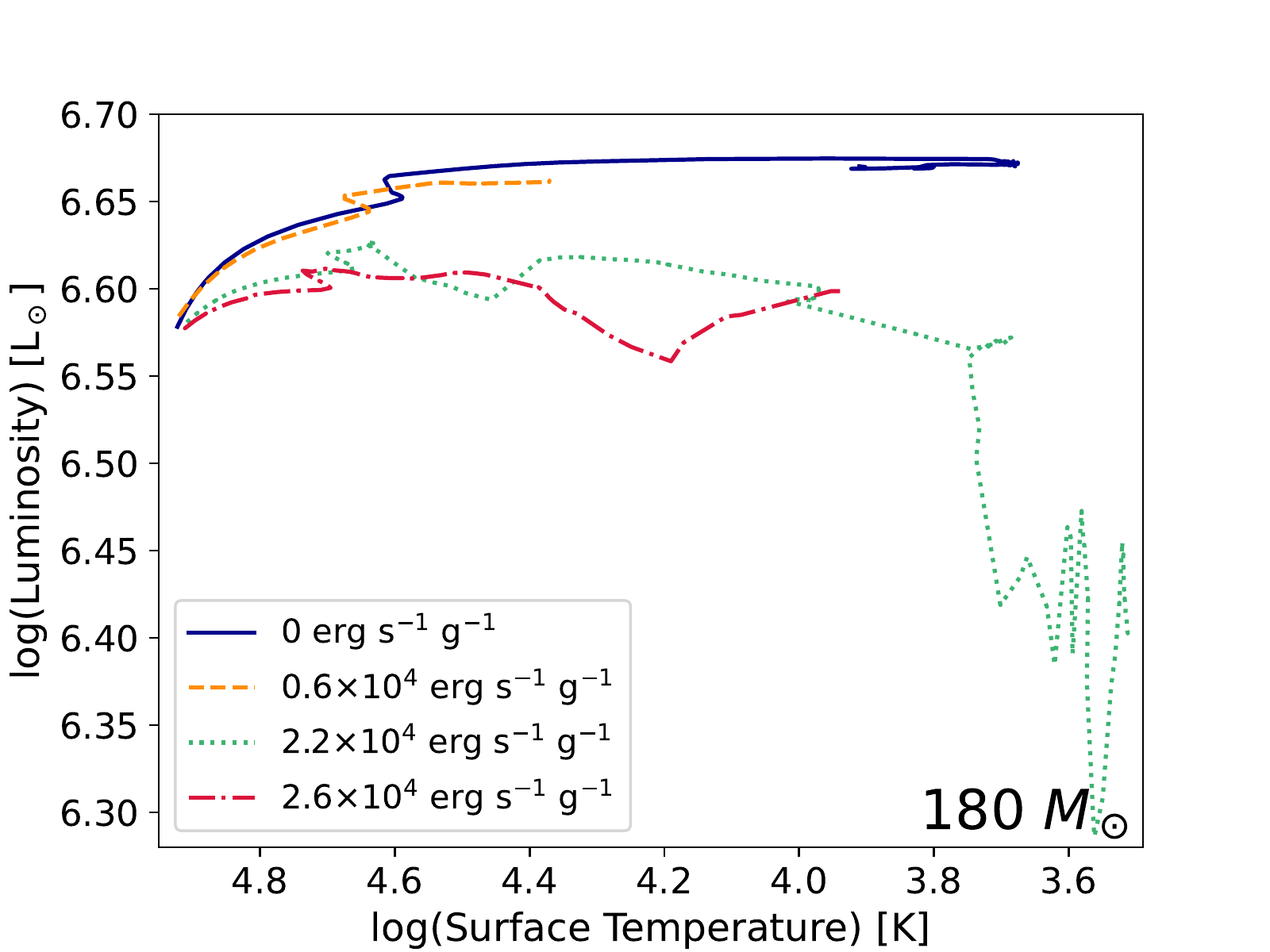}
    \includegraphics[width=0.49\linewidth]{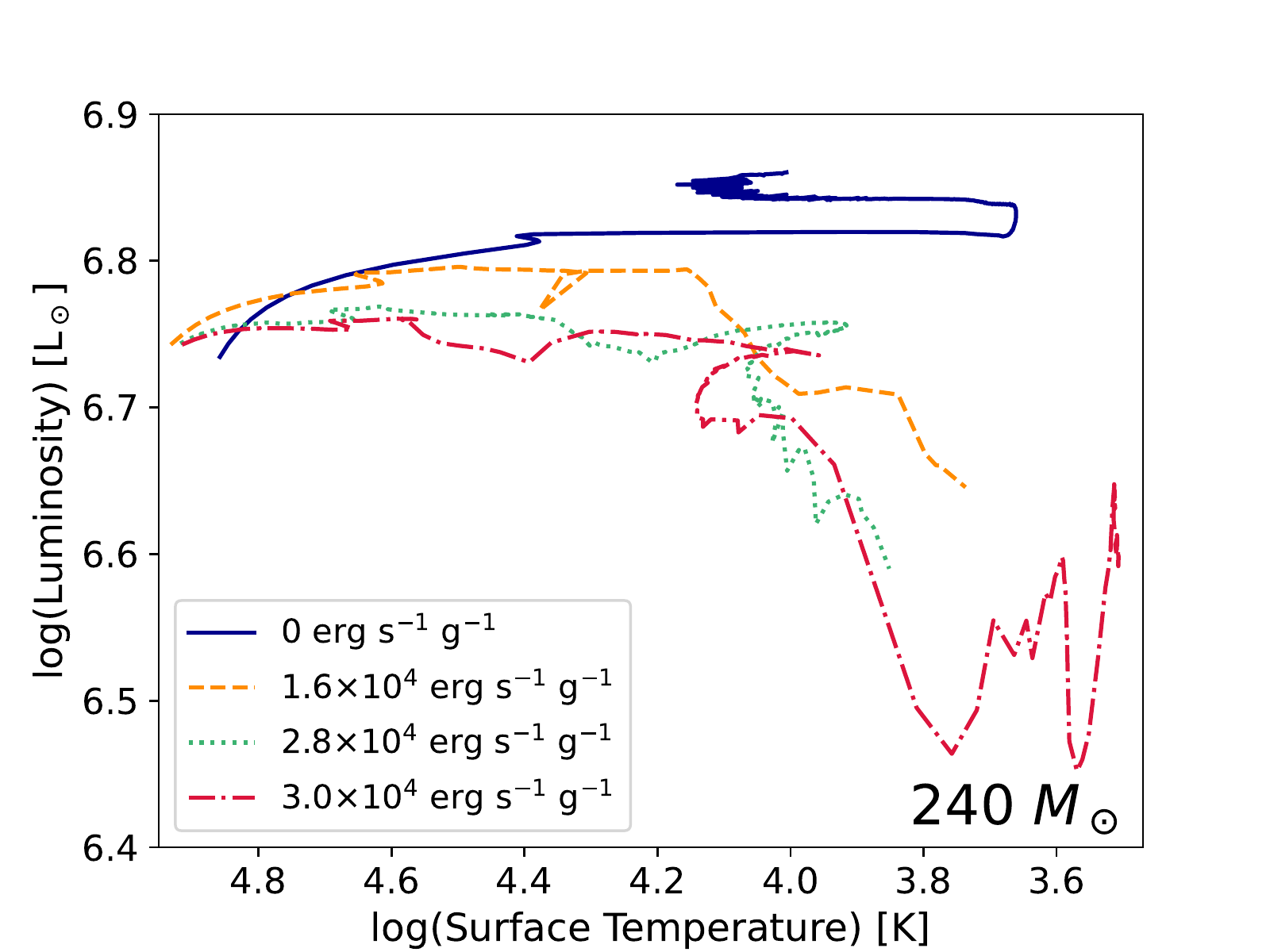}
    \includegraphics[width=0.49\linewidth]{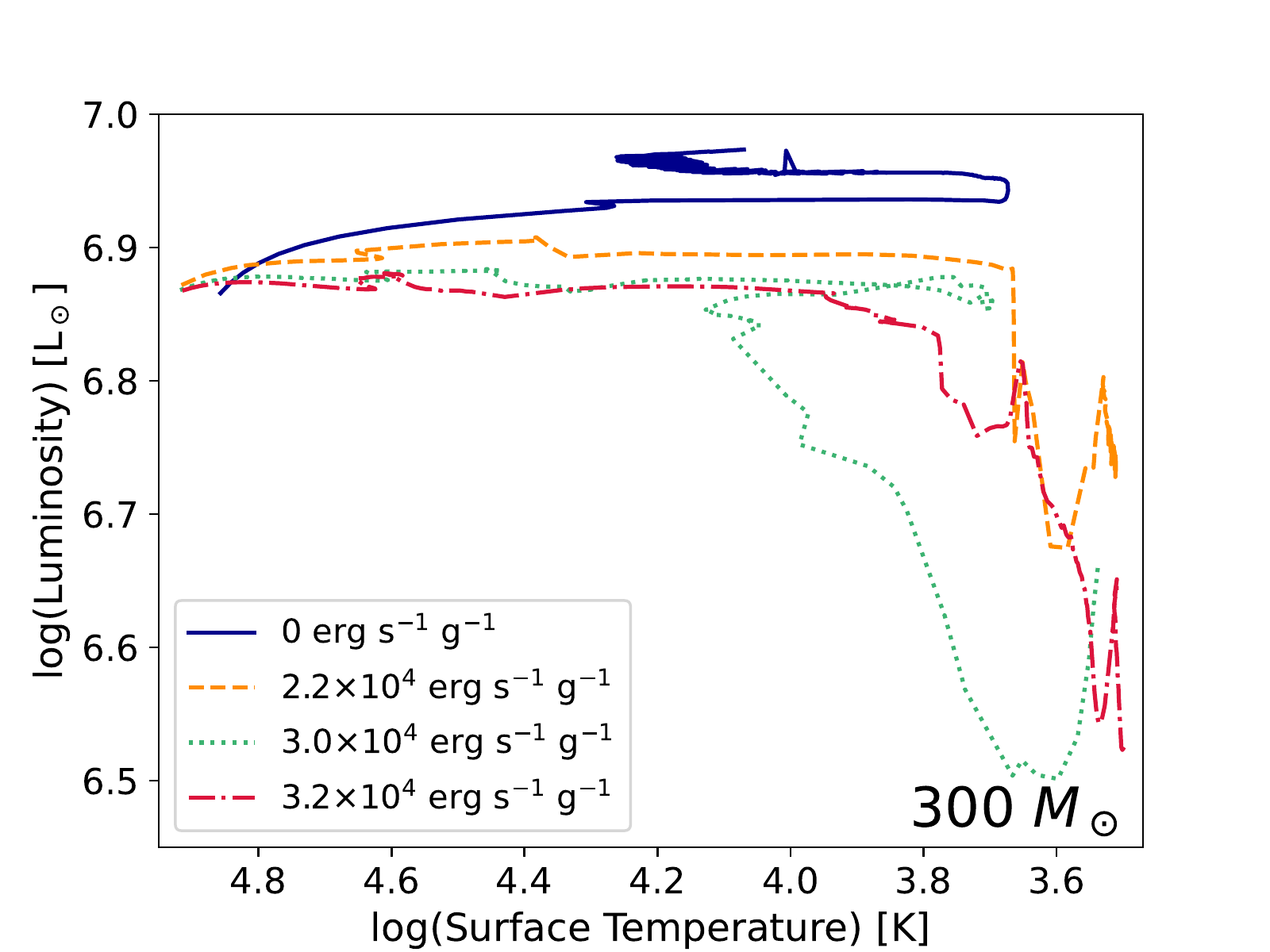}
    \caption{Hertzprung Russell Diagrams for stars with and without Non-Nuclear Energy (NNE) added. Each panel shows the evolution of stars of a different mass from the beginning of main sequence to approximately 99.9\% of the total stellar lifetime. In each panel, the blue curve corresponds to the evolution without NNE added, with other curves showing the evolution of stars with constant NNE added 
   as labeled. For comparison, the energy produced by nuclear fusion during the main sequence in the nuclear-only stars ranges from approximately $3 - 5\, \mathrm{erg\, s^{-1} \, g^{-1}}$. Increasing the amount of NNE causes the stellar surface to have a lower surface temperature, reflecting similarities with lower mass stars without NNE.}
    \label{fig:HR}
\end{figure}

More specifically, the evolution of stars with NNE shares some features of the evolution of nuclear-only stars with the same mass and some features of the evolution of nuclear-only stars with a lower mass. At the beginning of the main sequence, the addition of NNE affects the internal structure of the star, but does not substantially effect its observable properties, like luminosity or surface temperature. Consequently, when plotted on a Hertzprung-Russell (HR) diagram, stars will start at essentially the same location, regardless of whether they have NNE added or not. However, as stars evolve, the core that develops begins to take on a much larger role in determining the evolution. Therefore, stars with NNE will evolve differently from stars without, and furthermore, the more NNE is added the more the HR track will diverge from the nuclear-only case. Specifically, increasing the amount of NNE tends to cause the HR track to behave more like a lower mass star. In figure \ref{fig:HR}, we plot the HR diagram of each star from main sequence to late in the star's lifetime. Each subplot shows stars with the same mass but different amounts of NNE, and the behavior described above can be seen.

\begin{figure}
    \centering
    \includegraphics[width=0.49\linewidth]{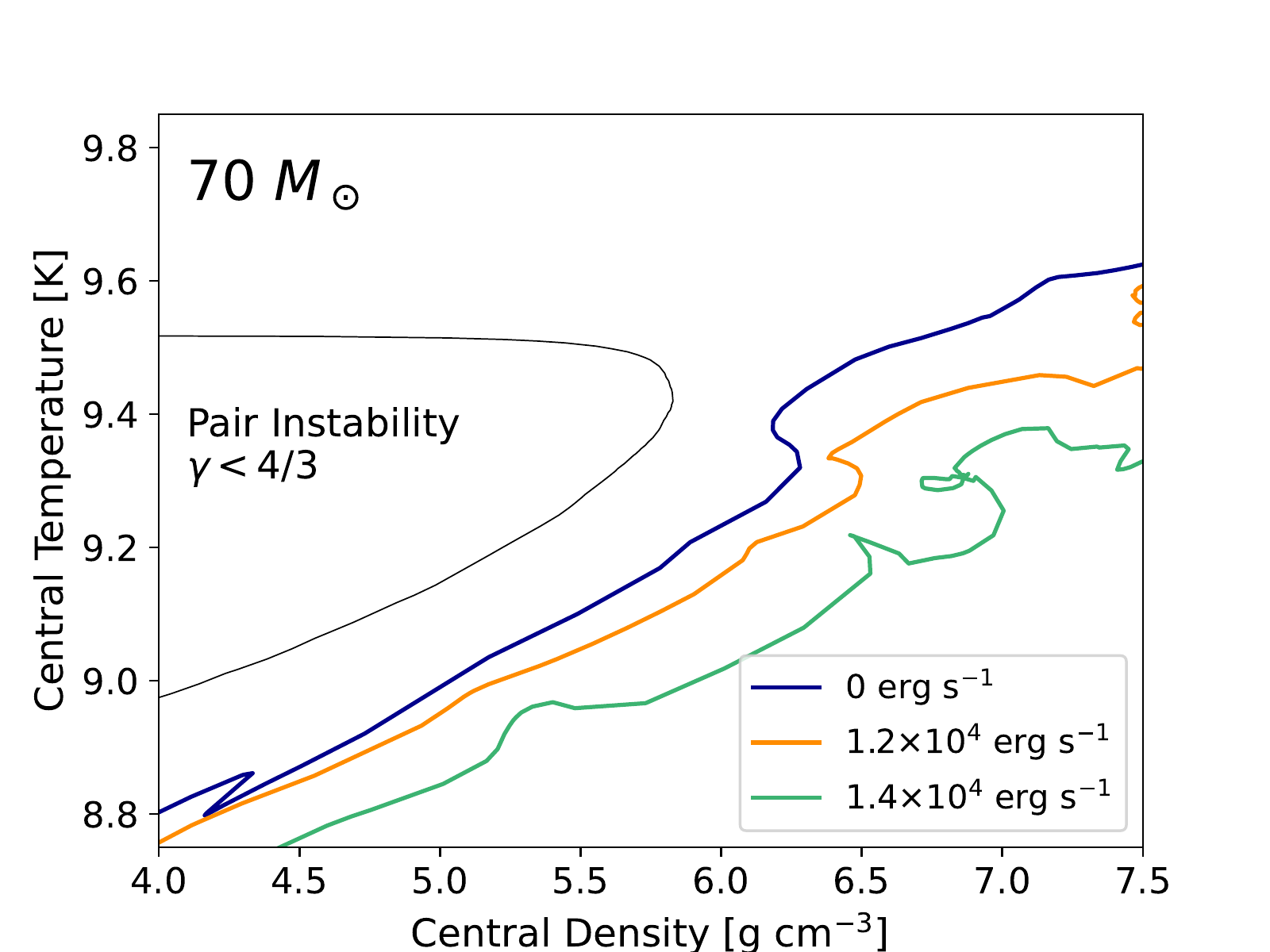}
    \includegraphics[width=0.49\linewidth]{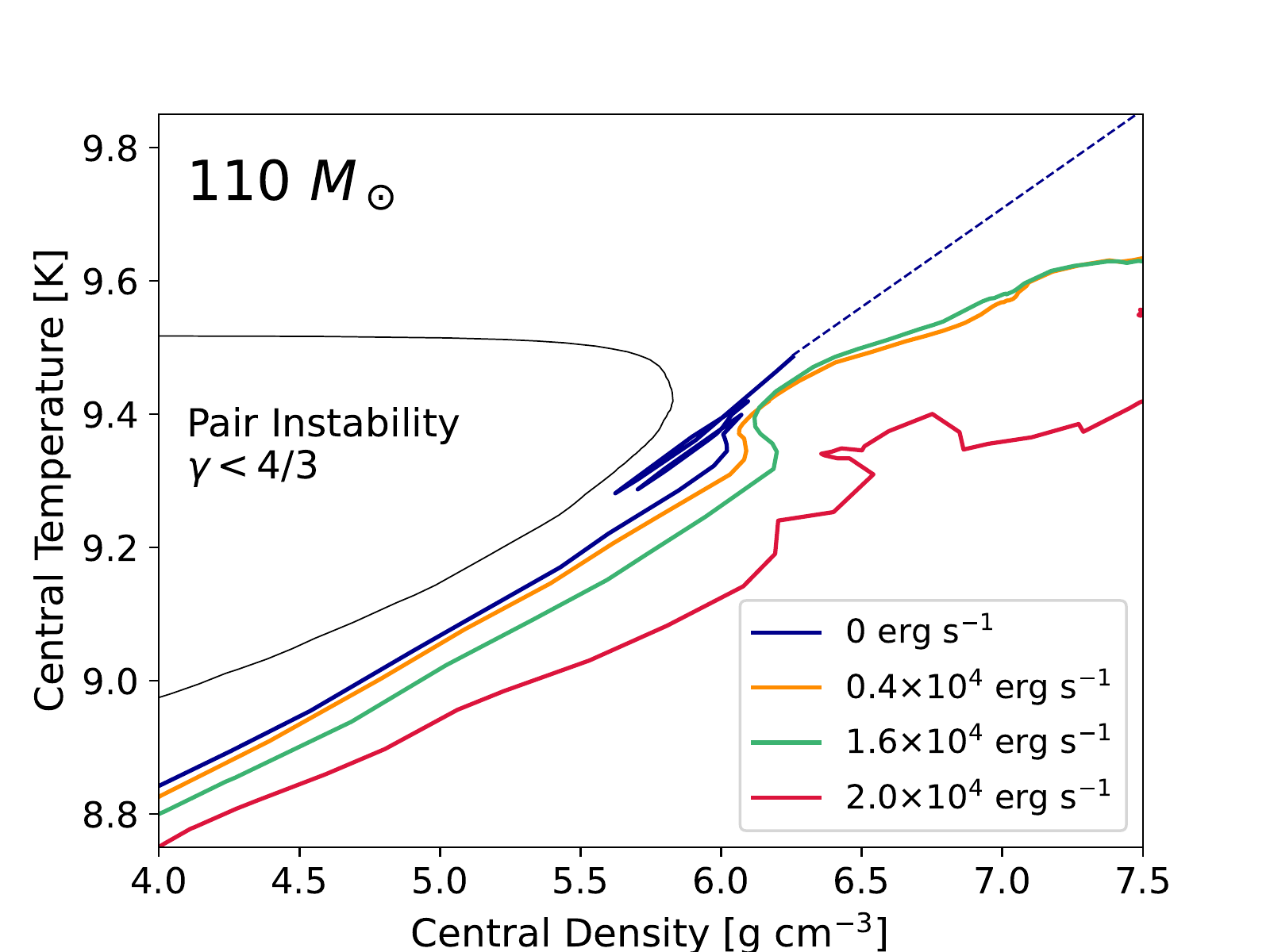}
    \includegraphics[width=0.49\linewidth]{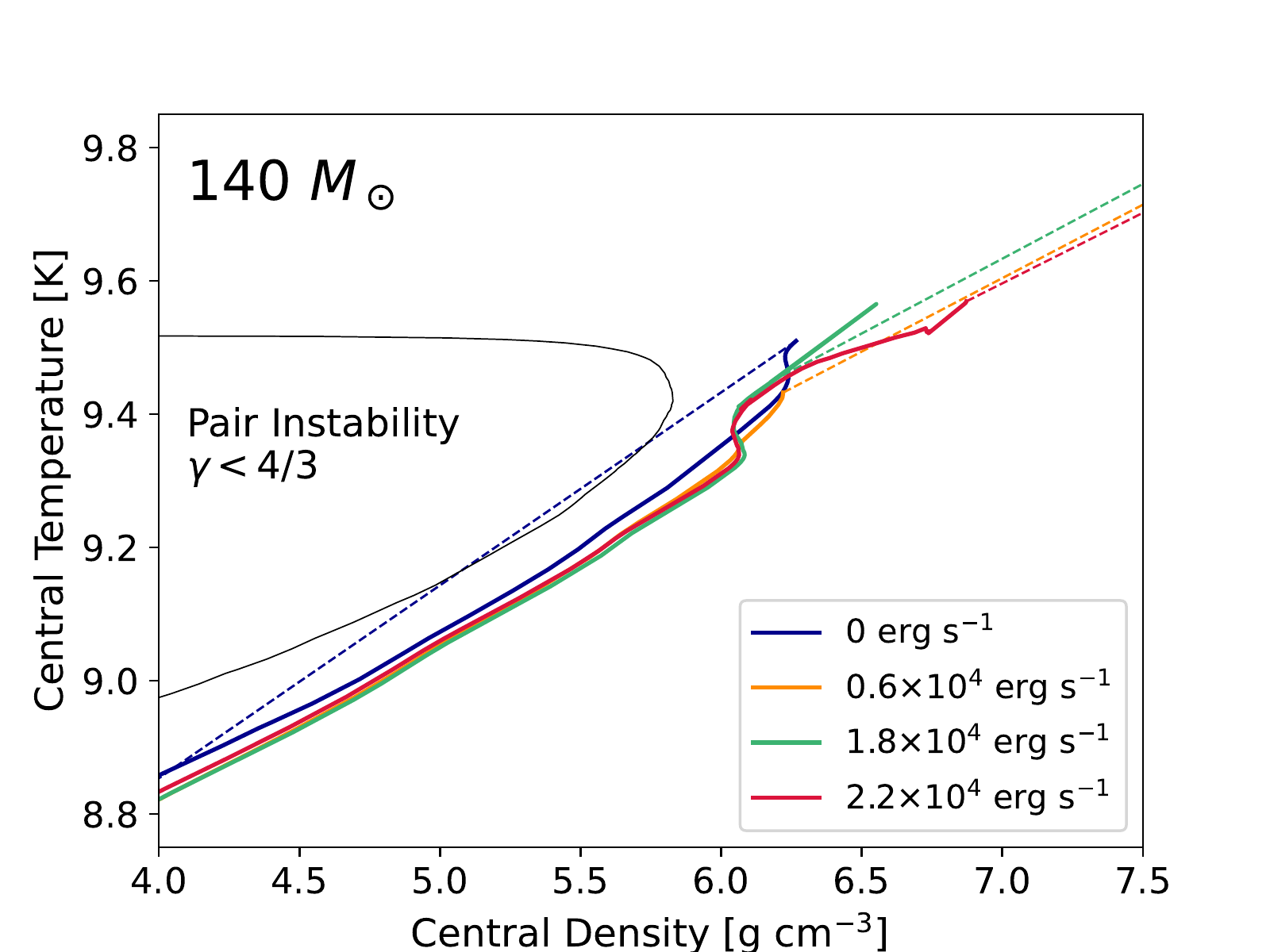}
    \includegraphics[width=0.49\linewidth]{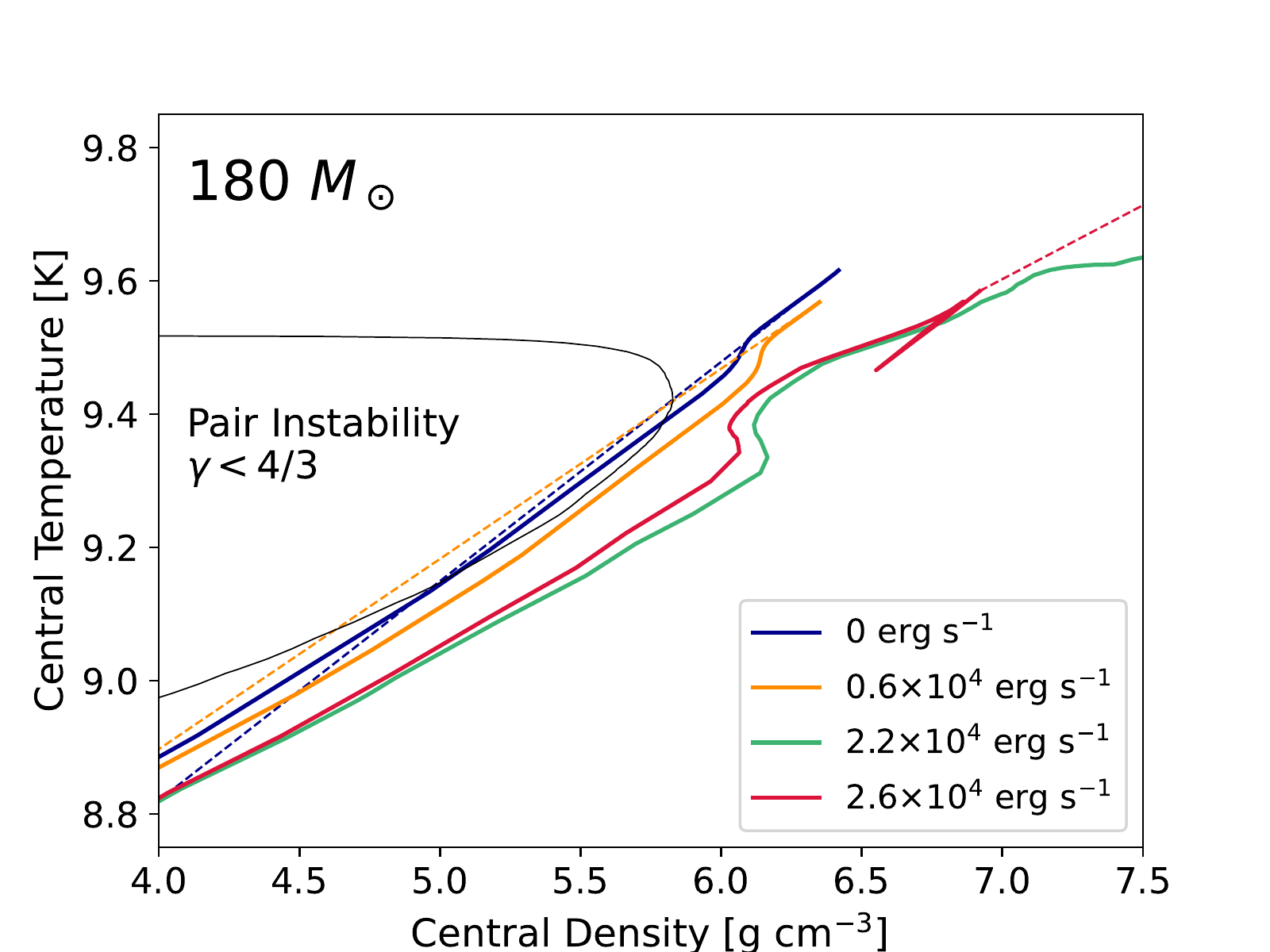}
    \includegraphics[width=0.49\linewidth]{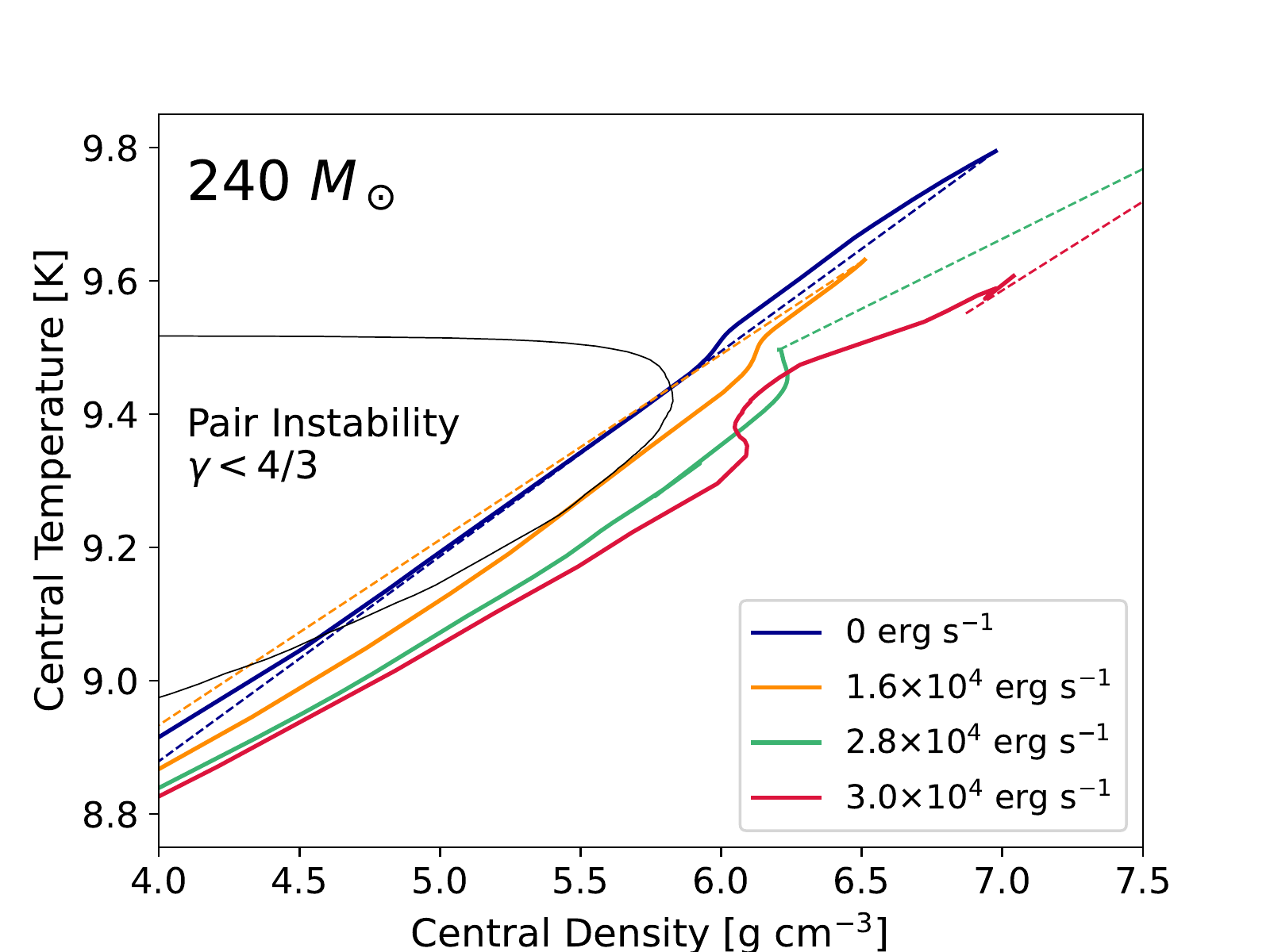}
    \includegraphics[width=0.49\linewidth]{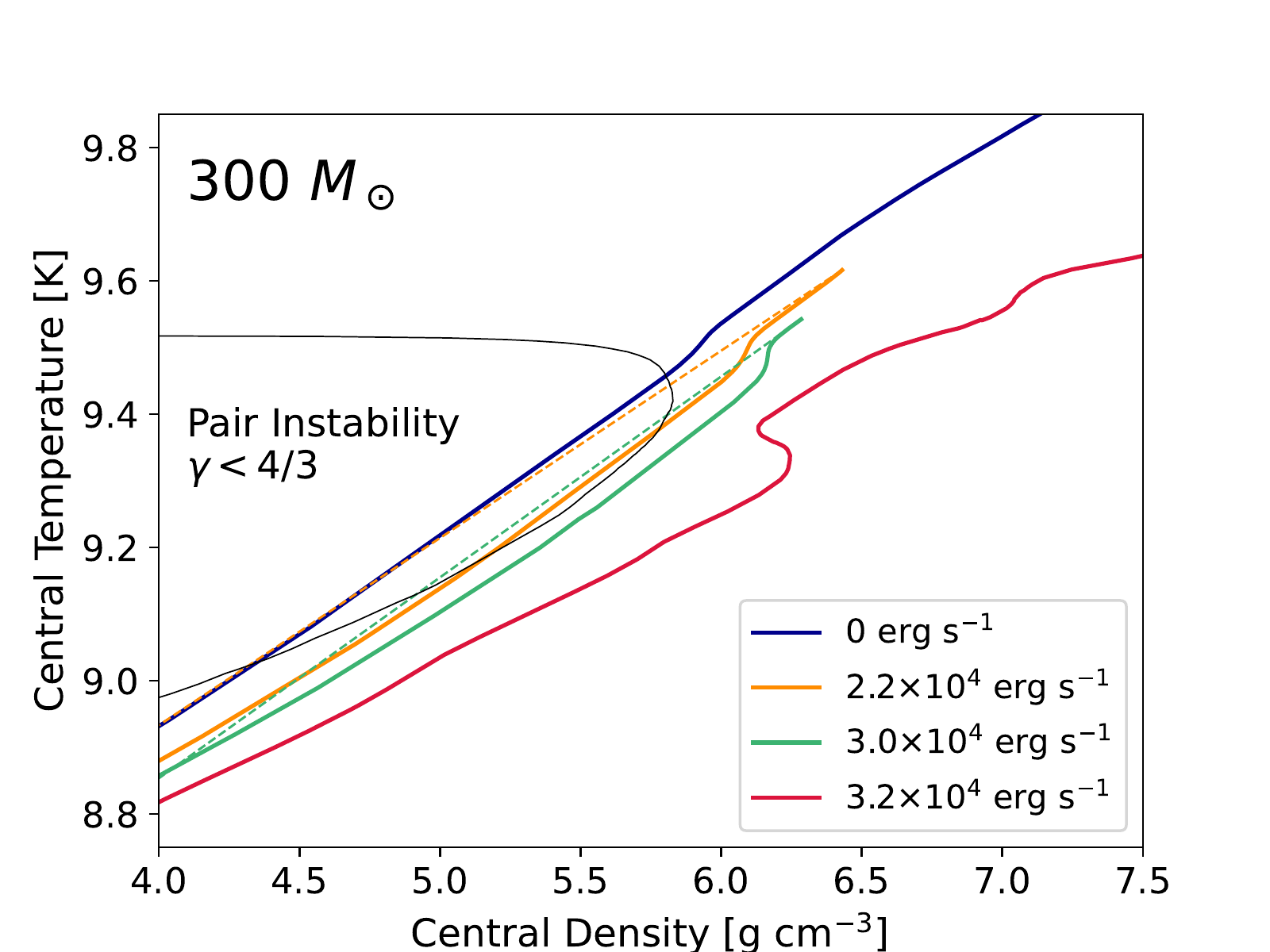}
    \caption{Temperature-Density plots for stars with and without Non-Nuclear Energy (NNE) added. Each panel shows evolution of stars with the same mass but different amounts of NNE as labeled in the legend. Within the half-elliptical thin black lines is the region where pair instability occurs. Tracks that avoid this region evolve to core collapse supernova progenitors (and eventually black holes), while those that enter the pair instability region evolve to pair instability supernovae (that leave no remnant). Between these cases, stars undergo pulsations in which material is ejected from the star. To reduce clutter, all portions of the evolution of a star after the first collapse are replaced by a dashed line to the final density and temperature.}
    \label{fig:Trho}
\end{figure}

In addition to HR diagrams, we can see how the smaller core of stars which have NNE added affects the central density and central temperature in the star. Figure \ref{fig:Trho} plots the stellar evolution of stars with and without NNE, and examines which ones enter the PISN region inside the black solid lines.  We will describe the behavior of the different stars in the next few paragraphs.

For all stars, the central temperature and central density both generally increase with time, and very roughly follow the evolution $\log T_c/ \log \rho_c \approx 1/3$, which is what is predicted by the simple Eddington solar model. However, more precise modeling of stellar evolution adds nuance to this idealized curve. In fact, low mass stars tend to have $\log T_c/ \log \rho_c \lesssim 1/3$ while higher mass stars can see $\log T_c/ \log \rho_c \gtrsim 1/3$. In addition to the changing slope, higher mass stars tend to have higher central temperatures at any given phase of evolution than their lower mass counterparts. These differences becomes particularly noticeable when discussing pair instability. 

Pair instability occurs when the production of electrons and positrons reduces the pressure sufficiently enough for $\gamma= \partial \log P/ \partial\log \rho < 4/3$. At this value, a fluid is no longer pressure-supported and becomes unstable to collapse. For a given equation of state, this expression defines a fixed contour in the density-temperature plane. Then, different stars will either pass by or cross through this region as they evolve. For example, in this language, the temperature-density tracks traced by the central temperature and central density of high mass stars tends to be closer to this pair instability region than lower mass stars. Like in the HR diagram, we see again that NNE causes the tracks on these diagrams to shift to look more like those from lower mass stars powered by nuclear reactions.

In Figure~\ref{fig:Trho}, we plot the temperature-density tracks made by the centers of each of the stars we considered, along with the pair instability region. When a given stellar track crosses into the unstable region, this tends to lead to a PISN. When tracks pass close to, but do not actually pass into the unstable region, pair instability does not happen in the center of the star but likely does happen somewhere off-center, which tends to lead to PPISNe. Tracks that avoid the pair instability region by a sufficiently wide margin instead evolve to a separate instability region which extends in a band across all densities off the top of the plotted temperature range and which leads to a core collapse supernova. And finally, some tracks enter the pair instability region, but then leave before the star has the chance to fully collapse. In these cases, the corresponding stars continue evolving to undergo direct collapse to form a black hole, which we treat as a special case of core collapse supernova.

\subsection{Core-Envelope Interface}

The evolution of nuclear-only stars is often modeled only considering the stellar core, but the addition of NNE requires that we consider the entire star, both core and envelope. In particular, the addition of NNE to a star can significantly alter the stellar structure and evolution at the boundary between the stellar core and envelope. There are three main causes for these changes: an increase in hydrogen content in the envelope, a greater ability to mix across boundaries, and a lower amount of energy provided by nuclear sources in the core.

During the main sequence, stars with NNE fuse less hydrogen into helium than stars of the same mass without NNE. On one hand, this means that during formation, stars with NNE do not contract as much before enough energy is produced to support the star. As a consequence, stars with NNE have lower temperature and density than their nuclear-only counterparts. This leads stars with NNE to convert smaller amounts of hydrogen into helium, and as the star evolves, a smaller core and a larger envelope develop. Furthermore, as the star evolves, it contracts and density and temperature increase everywhere throughout the star. In stars with NNE however, the increase in temperature and density of the hydrogen just outside the core of the star can become high enough to cause significant hydrogen fuel in a hydrogen-burning shell at the envelope-core boundary. While hydrogen-burning shells also exist in nuclear-only stars, the smaller envelopes of nuclear-only stars typically prevents the shells from playing a major role in the star's evolution. In stars with NNE, however, full support for the star can come from the hydrogen shell burning and the NNE that is added, at least until hydrogen in the envelope is sufficiently depleted. Furthermore, in many cases, the shell-burning causes the outer portions of the envelope to expand significantly, inducing a sharp drop in density at the boundary between core and envelope.

In addition to the increased energy achievable through hydrogen-shell burning, the addition of NNE also causes changes to the convection within a star, and these changes can lead to mixing across the core-envelope interface. 
In general, whether or not a region of the star is undergoing convection can be determined from the Ledoux criterion\cite{ledoux1947, kippenhahn}:
\begin{equation}
\frac{d \ln T}{d \ln P}_{rad} < \frac{d \ln T}{d \ln P}_{ad} + \frac{\varphi}{\delta}\frac{d \ln \mu}{d \ln P}_s,
\end{equation}
where the subscript $rad$ signifies that the differential is to be taken as if the star were only transmitting energy through radiation, and the subscript $ad$ signifies that the differential is to be taken assuming adiabatic processes. Similarly, the subscript $s$ corresponds to taking the differential at a fixed entropy. Finally, $\varphi = \frac{d \ln \rho}{d\ln \mu}$ and $\delta = - \frac{d \ln T}{d \ln \mu}$ are taken at fixed pressure and either temperature or density, respectively.
If a region of a star satisfies the Ledoux criterion, it is stable against convection and transmits energy primarily through radiation. Conversely, those regions of the star where this condition is not satisfied undergo convection and experience chemical mixing. Between convectively mixed regions there tend to exist boundaries which resist convective mixing because of the $\frac{d \ln \mu}{d \ln P}_s$ term, establishing relatively stable regions of uniform chemical composition.

Introducing an NNE source can weaken these natural boundaries by causing the other terms in the Ledoux criterion to change. Heuristically, we can relate $\frac{d \ln T}{d \ln P}_{rad}\propto P/T^4$ and $\frac{d \ln T}{d \ln P}_ad \propto P/T\rho$. The addition of NNE to a star causes its evolution to include some features of lower mass stars. In general, stars with a lower mass have a lower central temperature and higher central density, suggesting that adding NNE could change the density and temperature profiles in a nontrivial way. And indeed, this is what we see in our simulations. Increasing the amount of NNE in a star allows convective mixing to happen across the chemical boundaries more easily. Simultaneously, the strong hydrogen-shell burning discussed above can further affect the temperature profile near the core-envelope interface. Consequently, in several of our simulated stars with sufficiently large amounts of NNE, we see the inner edge of the envelope move to lower masses during the hydrogen-shell burning phase, implying that there is mixing between the envelope and core. Furthermore, substantial fractions of metals (up to a mass fraction of 10\%) especially carbon, nitrogen, and oxygen can migrate outward and eventually reach the surface of the star through convection in the envelope. As the shell burning continues, it gradually reduces the mass of the core slightly, which can further impact the stellar evolution.

The inward migration of the core-envelope boundary has a second major effect when the star nears pair instability. Late in its evolution, a massive nuclear-only star has an onion-skin structure, with multiple shells undergoing different types of shell-burning. At the main envelope-core interface, there is hydrogen burning, and inward of this there are shells of helium burning, carbon burning and so on. As a result, there are multiple relatively independent shells of different chemical composition. During pair instability, these shells can play a role in preventing the outward propagation of pair-instability-induced density waves. That is, small amounts of pair instability may cause the formation of a density wave in the star, but it typically stalls in the core of the star at boundaries between different shells before propagating to the low density envelope. Stars near 100 $M_\odot$ that ultimately undergo core collapse but have masses close to those that undergo PPISN typically experience this phenomenon. Conversely, when the core-envelope interface migrates inward in stars with NNE, these shells are prevented from forming or eliminated if they have already begun to form.

In turn, when stars with NNE undergo small pair instability density waves, those waves tend to propagate to the core-envelope interface without completely stalling out. In addition, because of the hydrogen-shell burning, the core-envelope interface also coincides with a very sharp drop in density. When pair instability density waves reach this interface, they can be reflected back into the core of the star. The increased density in the center of the star can trigger another burst of oxygen-burning, causing a larger wave to bounce outward. Effectively, the core of stars with NNE added exhibit forced damped oscillation that is not seen in the nuclear-only stars we look at, or is seen only as a minor effect. Eventually, over multiple periods, the energy of the wave increases until it is greater than the escape velocity of the star at the interface. We designate the moment the escape velocity is surpassed as an oscillatory burst event.
Unfortunately, how the oscillatory burst event affects the final core collapse supernova is beyond the scope of this paper. In fact, in many cases, this oscillatory burst event occurs only seconds before the onset of a core-collapse supernova, with the longest period between oscillatory burst event and supernova that we simulated only hours.

For the purposes of this study, we treat stars that experience this oscillatory burst event as core-collapse supernovae, not as PPISN. While there are some similarities, PPISN typically occur when a larger density wave escapes the star without oscillating within the core. Furthermore, in at least one simulated evolution, the mass of the star below the escape velocity increased after the oscillatory burst event. That is, the material ejected by the oscillatory burst event does not transfer much energy to its surroundings, and the envelope through which it travels may remain bound to the star. A large uncertainty remains in how core collapse would affect the stellar material that undergoes this oscillatory burst event, but unfortunately that is beyond the scope of this work. However, if these oscillatory burst events do occur, they may leave signatures in the supernova light curves that could be detectable. 

\subsection{Predicting final state}

One of our primary goals in this project is to predict the outcome of a star's evolution with reasonable accuracy without the need to individually model every combination of stellar mass and NNE added. To that end, we classify combinations of mass and NNE into three categories based on the type of supernova a star with that initial mass and NNE would evolve to: core-collapse, PISN, and PPISN. For combinations whose evolution we simulated, we make this classification simply by looking at the simulations. In most cases, we can classify the evolution solely from the final mass of the evolution ($M_{fin}=0$ for PISN, $M_{fin} \lesssim 1/2 M_{in}$ for PPISN, and $M_{fin} \gtrsim 1/2 M_{in}$ for core collapse). However, as discussed above, we classify those combinations of stellar mass and NNE that lead to a stellar evolution with what we call an oscillatory burst event as leading to a core-collapse supernova, even though it mimics a PPISN based only on the final mass.

We then extend the classifications of these individual simulations to predict which evolutionary outcome stars with any combination of stellar mass (between 70 and 300 $M_\odot$) and NNE achieve. First, we consider separately each of the six masses we simulate. For each mass, we identify an amount of NNE that roughly corresponds to the boundary between two different evolutionary outcomes. For example, to determine a boundary NNE between PISN and PPISN, we use a value between the highest amount of NNE that leads to PISN and the lowest amount of NNE that leads to PPISN. In most cases, this value that is chosen is simply the midpoint between the two amounts of NNE (that lead to PISN and PPISN). In some cases, however, we adjust this value based on either features of one or both of the evolutions on either side of the boundary, or evolutions which did not fully complete but have behaviors that suggest one evolution or the other. 
For example, for the 180 $M_\odot$ boundary between PPISN and core-collapse, this approach was used. Once a boundary NNE was determined for each of the stellar masses we consider, we linearly interpolate between these boundary points to estimate the boundary NNE at stellar masses between the masses we simulated.

\begin{figure}
    \centering
    \includegraphics[width=0.9\linewidth]{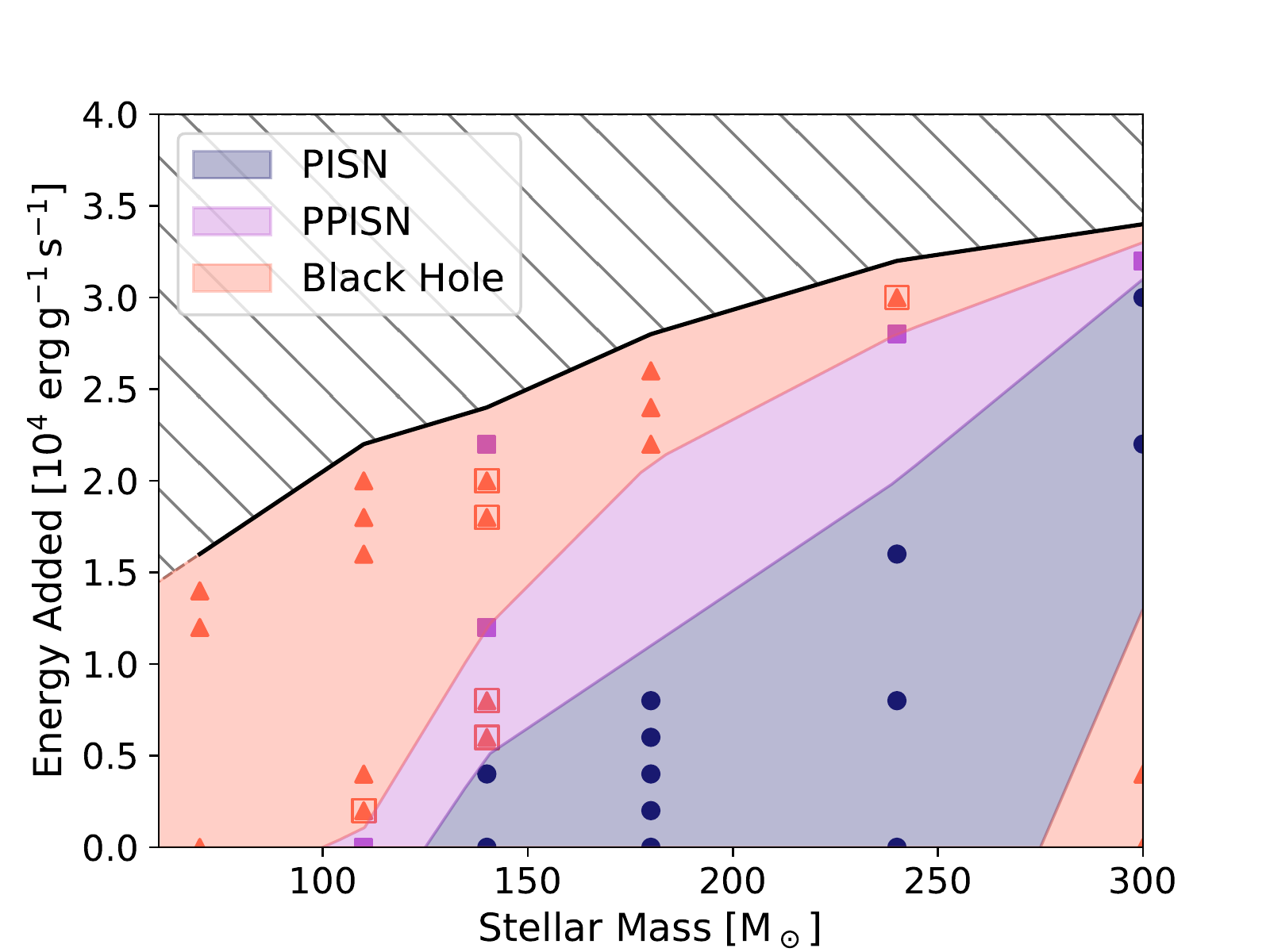}
    \caption{Final evolutionary outcomes for stars of the six initial masses considered in this paper, when different amounts of NNE are added. Each simulation is represented by a point, located based on the initial mass of the star and the NNE density added to the star. Red triangles correspond to simulations that evolve to core collapse supernova (and eventually black holes), blue circles correspond to simulations that evolve to PISN (and leave no remnant), and magenta squares correspond to simulations that evolve to PPISN (which leave behind a small remnant, significantly lower mass than the initial stellar mass). 
    Stars that evolve to core collapse but experience an oscillatory burst event are identified as red triangles within squares. Contours of the same colors are constructed to estimate the evolutionary behavior of stars that would exist between these points (see main text for details of how this estimation is performed). Simulations of stars in the hashed region do not initially converge to form stable stars. Notably, for any initial stellar mass, there is an amount of NNE that can be added to produce stars which avoid pair instability and evolve to core-collapse supernovae and eventually black holes.}
    \label{fig:summary1}
\end{figure}

\begin{figure}
    \centering
    \includegraphics[width=0.9\linewidth]{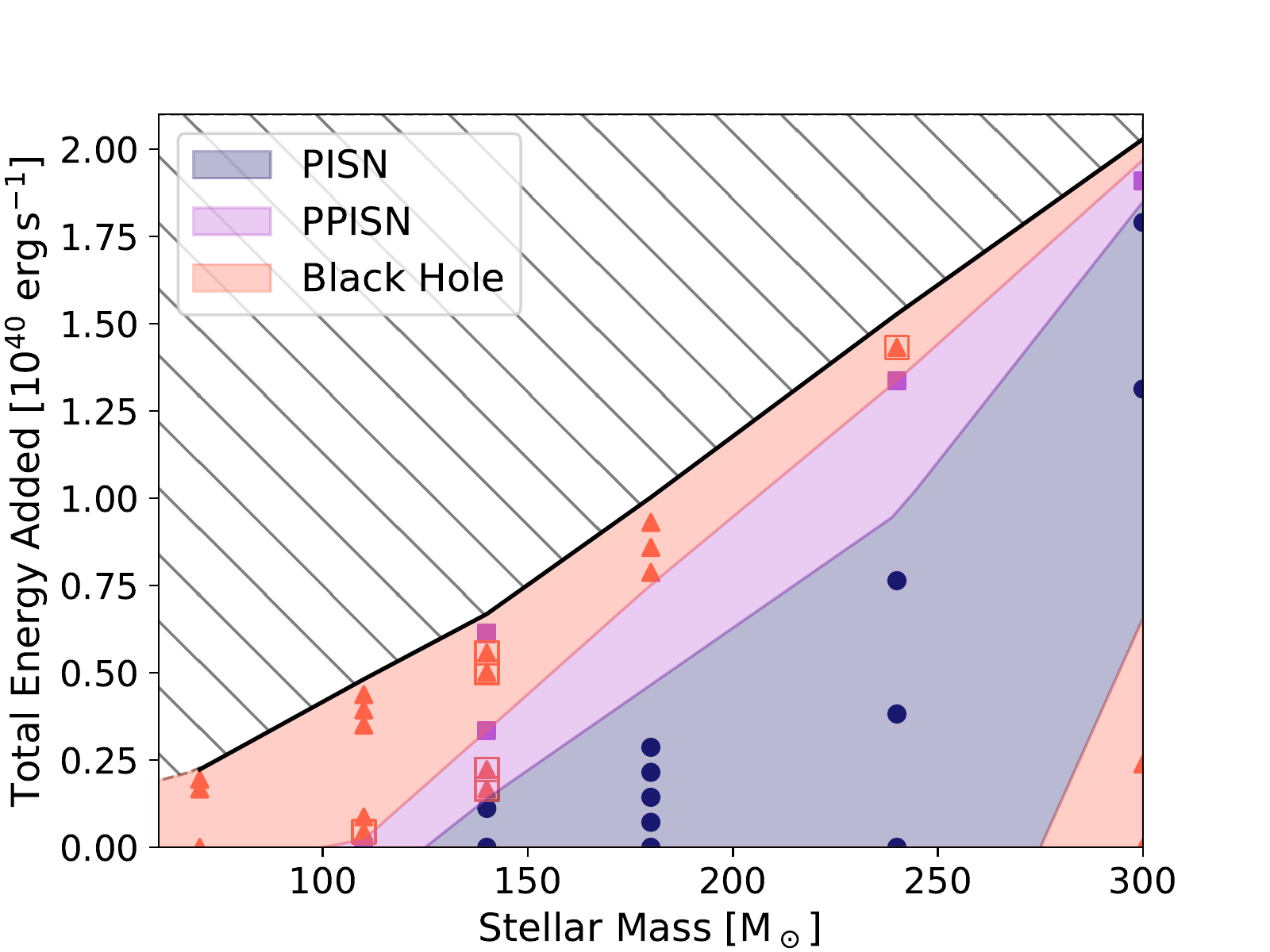}
    \caption{Same as figure \ref{fig:summary1}, but with simulations and contours plotted by the initial stellar mass and the total luminosity from NNE sources in a star. The luminosity (y-axis in this figure) is calculated as the energy production rate (y-axis in figure \ref{fig:summary1}) multiplied by the initial mass of the star (x-axis in either figure).}
    \label{fig:summary2}
\end{figure}

\begin{figure}
    \centering
    \includegraphics[width=0.9\linewidth]{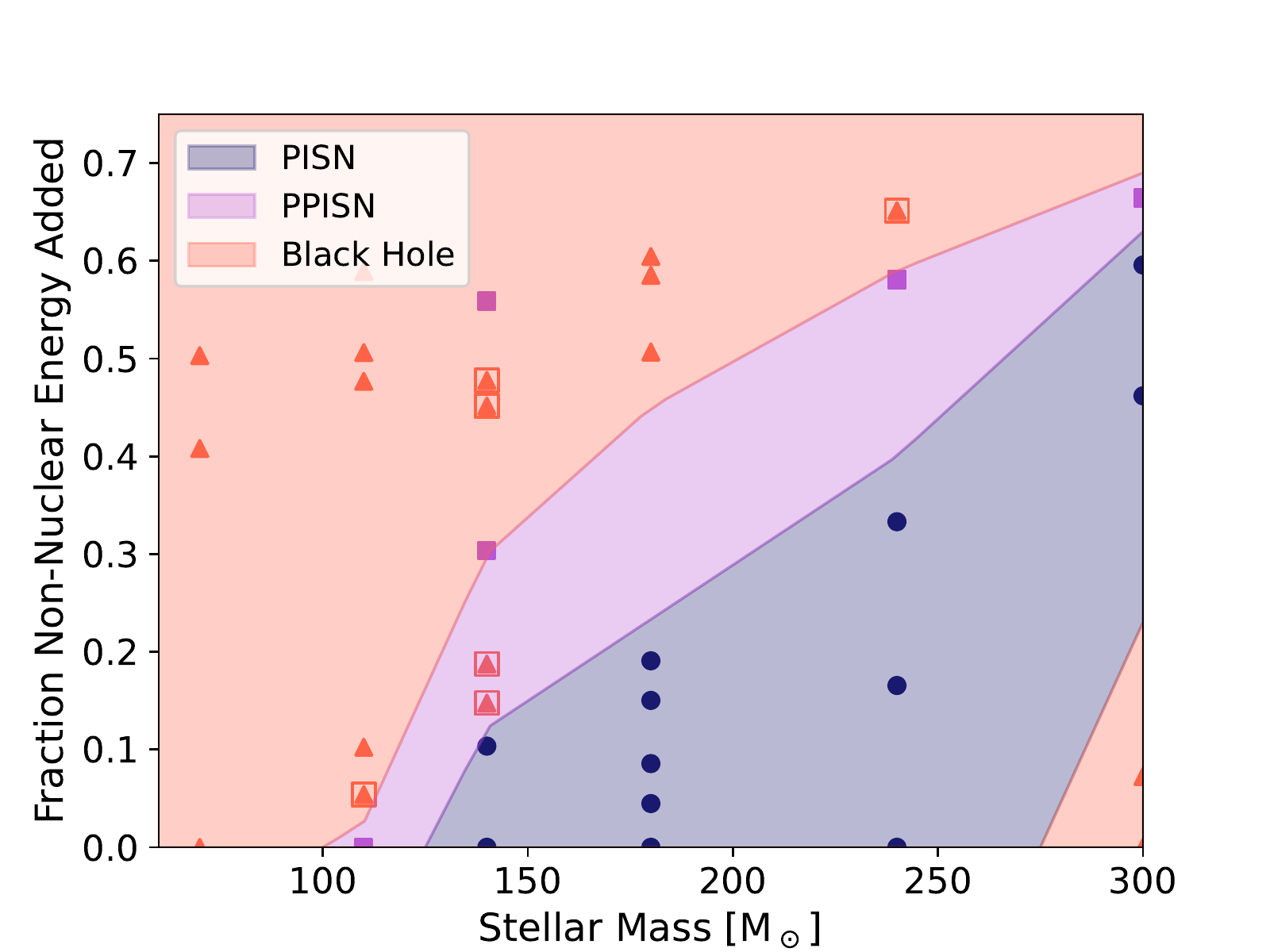}
    \caption{Same as figure \ref{fig:summary1}, but with simulations and contours plotted by the initial stellar mass and the fraction of main sequence luminosity contributed by NNE sources.}
    \label{fig:summary3}
\end{figure}

Using the boundaries established above, we can produce contour plots showing what combinations of mass and NNE we predict to lead to each evolutionary outcome. We present three such contour plots in figures~\ref{fig:summary1}, \ref{fig:summary2}, and \ref{fig:summary3}. While the simulations and boundaries are the same, we describe the amount of NNE added in three different ways. 
In the first case (figure~\ref{fig:summary1}), we consider the NNE in terms of the energy production rate per mass of stellar material. That is, the NNE here is described using the constant value added to equation~\ref{eqn:energyrate}. In the second case (figure~\ref{fig:summary2}), we consider the energy in terms of the integrated rate within the star, $\epsilon_{NNE} M_*$ for the initial mass of a star $M_*$. This is effectively the contribution to the stellar luminosity due to the NNE. Finally, in the third case (figure~\ref{fig:summary3}), we consider the energy in terms of the fraction of the luminosity contributed by NNE sources during the main sequence $\epsilon_{NNE} M_*/L_{MS}$, for the stellar luminosity during the main sequence $L_{MS}$.

One of the most notable results visible in these contour plots is that it is possible for stars with any initial mass to evolve to avoid pair instability and undergo core collapse supernovae if an appropriate amount of non-nuclear energy is added. In other words, as long as the physical mechanism behind the NNE is accessible to stars of any mass, black holes of any mass in the upper mass gap could be possible directly from stellar evolution. Admittedly, it is unlikely that the majority of stars will be injected with NNE, so black holes within the mass gap will be substantially rarer than black holes with masses either above or below the mass gap, but if NNE can be injected into stars, it should result in a non-zero floor in the number of black holes at any mass within the mass gap. We discuss this more below. At the same time, studies into the number of core collapse supernovae and their light curves could offer additional ways to probe the addition of NNE to stars. 

\section{Proposed Source of Non-Nuclear Energy}
\label{dark matter}

So far, we have discussed the impacts of adding NNE without discussing the source of that energy. All of the effects we discuss--the reduction in size of the stellar core, increase in the mass of the envelope, and the change in temperature and density profiles--depend predominantly on the fact that the energy we add is not nuclear in nature. Consequently, we expect that these effects will appear independently of the precise details of the NNE source. That said, it is useful to consider whether the necessary energy production rates could be produced in a physically realizable scenario. Here, we consider one example of such a mechanism and discuss the degree to which it must be fine-tuned to satisfy the necessary production rate.

If dark matter is made up of particles, 
there is the possibility of annihilation between dark matter particles and their anti-particles. The annihilation rate per volume is 
\begin{equation}
\Gamma_{ann} = n_{DM}^2 \langle\sigma v\rangle = \rho_{DM}^2 \langle\sigma v\rangle /m_{DM}^2 \, 
\end{equation}
  where $\rho_{DM}$ is the dark matter density, $m_{DM}$ is the mass of the dark matter particle, and $\langle\sigma v\rangle$ is the velocity-averaged annihilation cross section of the dark matter with its antiparticle.  Dark matter annihilation is a very potent heat source. Unlike fusion, where less than a percent of the nuclear mass is converted into energy for the star, most of the mass of the annihilating dark matter particles is converted into energy which can power the star. Dark matter annihilation can produce a range of standard model products, most of which quickly scatter and thermalize with the surrounding stellar material, and consequently raise the temperature of that stellar material. Therefore, the heating rate per volume of the star is given roughly by the product of $\Gamma_{ann} \times m_{DM} (1-f_\nu)$. 
  The factor $f_\nu$ corresponds to the branching ratio of dark matter particles annihilating into neutrinos and other non-interacting particles that can escape the star; all other standard model particles have a high enough scattering cross section with stellar material that they are unable to escape, and therefore deposit energy into the local stellar medium.
  The annihilation energy per unit volume deposited into the star is then 
 \cite{freese2016}
 \begin{equation}
    Q_{DM} = (1-f_\nu) \langle\sigma v\rangle \frac{{\rho_{DM}}^2}{ m_{DM}}.
\end{equation}

To perform numerical modeling of stellar structure and evolution, it is more natural to consider not the annihilation rate per unit volume but the annihilation rate per unit mass. To convert, we use the density of all of the matter (gas and dark) within the star as a conversion factor. However, because the mass of dark matter is small compared to the mass of gas in these stars, we can neglect the dark matter density and use only the gas density as a conversion factor. As such, the quantity we are interested in, the energy production per unit mass, is
\begin{equation}
    \epsilon_{DM} = (1-f_\nu) \langle\sigma v\rangle \frac{{\rho_{DM}}^2}{ m_{DM} \rho_H}
    \label{eq:dmeps}
\end{equation}
where $\rho_H$ is the density of the gas in the star.

Dark matter annihilation  is predicted by several models for dark matter. For example, the annihilation of Weakly Interacting Massive Particles (WIMPs) is a common target for dark matter searches: in direct detection experiments, at the Large Hadron Collider, and indirect searches for annihilation products.  The annihilation energy from WIMPs could also power  dark stars (e.g. \cite{spolyar2008, freese2008, spolyar2009, Freese_2010, freese2016}), a proposal for the first stars to form in the Universe. These dark stars would be fully supported by annihilation of dark matter, as opposed to the stars we consider throughout this paper, which are only partially supported by NNE. Because of the dark matter annihilation as an energy source, dark stars are predicted to have low surface temperatures, which would reduce the amount of feedback they provide on the surrounding gas, and in turn allow dark stars to reach high masses $\sim 10^7 M_\odot$ and luminosities $\sim 10^{10} L_\odot$ and be detectable by the James Webb Space Telescope. However, WIMPs are not the only possible dark matter model which would predict annihilation which could partially or fully support a star. Self-interacting dark matter, made up of a twin-Higgs model, has been studied in work like \cite{Wu_2022}. Freeze-in dark matter models may also be capable of this sort of annihilation.

For concreteness, throughout the remainder of this paper, when we consider dark matter as an explanation for the NNE source, we consider WIMP dark matter. Under the WIMP paradigm, some constraints on the mass and cross section of the dark matter can be placed in order to see evolutionary changes like those we describe. (In other dark matter models, comparable constraints may be placed). For WIMPs, the annihilation cross section is set by the weak interaction cross-section, implying $\langle\sigma v\rangle \approx 3 \times 10^{-26} \mathrm{cm^3/,s^{-1}}$. Furthermore, we can exclude dark matter with masses outside the GeV- 10 TeV range for two reasons. Higher mass particles are ruled out by unitary constraints, while lower mass particles evaporate from the star so that the density of dark matter is never high enough to significantly change the evolution of the star. This occurs because interaction between the dark matter and the high temperature environment in the core of the star leads low mass dark matter to develop a velocity distribution with a significant fraction above the escape velocity. This bound has been estimated by \cite{Garani:2021feo} to be approximately 1 GeV.

Finally, the dark matter density is a quantity heavily influenced by the astrophysical environment in which the star evolves. For example, while the average density of dark matter throughout the universe is $\Omega_m \rho_{crit} \approx 8.5 \times 10^{-30} \mathrm{g /, cm^{-3}}$\cite{Planck2020}, the local density can be much higher or lower. For our purposes, we need the density to be much higher, and we can achieve this in multiple ways. For example, dark matter scattering off of baryons in stars offers a way to increase the density of dark matter in the core of stars\cite{PressSpergel1985}. In this process, dark matter scattering off of stellar matter can reduce the energy of the dark matter particle so that it becomes bound. Subsequent scatterings could then cause the dark matter to accumulate in the core of the star, and the density of that accumulated dark matter will depend on both the scattering and annihilation cross sections. If the scattering cross section is high and the annihilation cross section low, the density could be significantly higher than the ambient dark matter. In addition, through the process of adiabatic contraction\cite{blumenthal1986}, as baryonic matter contracts, because of a conserved action of the combined baryon-dark matter system, dark matter contracts as well. As a consequence, the central density of the dark matter halo also increases. Either dark matter capture or adiabatic contraction may be further boosted if the ambient dark matter density is higher. For example, it has been hypothesized that the dark matter around black holes can form very dense spikes, with densities potentially as high as $10^5 \mathrm{g \, cm^{-3}}$\cite{Gondolo_1999, Eroshenko2016}. If a star evolves within one of those spikes, capture of dark matter by the star could lead to very high densities of dark matter in the stellar core, even with a relatively low scattering cross section. 

\begin{figure}
    \centering
    \includegraphics{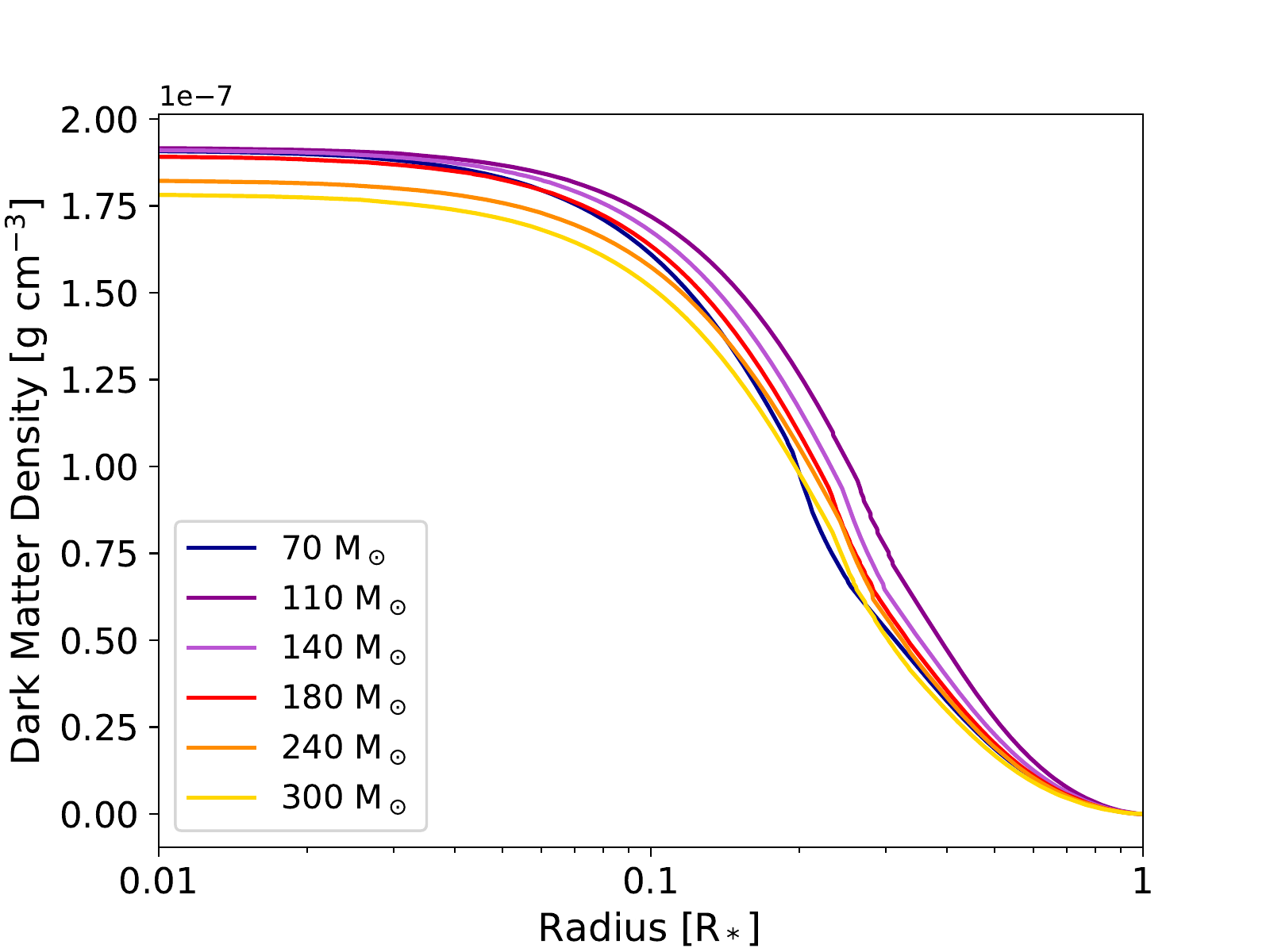}
    \caption{Density profile of WIMP dark matter necessary to produce approximately 60\% of the energy to support a star during its main sequence, as a function of the stellar radius, $R_*$. These profiles are derived from equation~\ref{eqn:density} in the main text. Gas density profiles are taken from stars with NNE that contributes closest to 60\% of the energy for each stellar mass. The NNE rate necessary to achieve this fraction is likewise estimated from the energy rates necessary to achieve fractions close to 60\%. If WIMP densities reach $\sim 2\times10^{-7} \mathrm{g\,cm^{-3}}$ during the main sequence, we can see the types of evolutionary changes we discuss throughout the paper, for stars of any of the masses we consider.}
    \label{fig:dmdens}
\end{figure}

With these mechanisms in mind, we can turn to the dark matter densities necessary to achieve the amount of energy we consider for the effects above. We can begin by rewriting equation~\ref{eq:dmeps} as
\begin{equation}
    \rho_{DM} = \sqrt{\frac{1}{1-f_\nu}\epsilon_{DM} \rho_H \frac{m_{DM}}{\langle\sigma v\rangle}}.
    \label{eqn:density}
\end{equation}
In Figure~\ref{fig:dmdens}, we plot the dark matter densities necessary in a star for 60\% of the energy in the star to come from the NNE. In this figure, we assume a dark matter particle mass of $1 \mathrm{GeV}$, an annihilation cross-section of $\langle\sigma v\rangle \approx 3.24 \times 10^{-26} \mathrm{cm^3\,s^{-1}}$, and a negligible fraction escaping as neutrinos. For the stellar density $\rho_H$, we use the density of stars with NNE that accounts for close to 60\% the total energy during the main sequence. Shown in this way, we can expect that densities on the order of $\rho_{DM} \sim \mathcal{O} (10^{-7} \mathrm{g \, cm^{-3}})$ would be necessary in order to see the effects we describe above, like the formation of mass-gap-filling black holes. As discussed above, while it is rare for dark matter to reach such high densities, there are scenarios where we can expect to see them.

\section{Black Hole Initial Mass Function}
\label{BHIMF}

In the not-too-distant future, next generation gravitational wave observatories are predicted to observe hundreds of thousands of black hole mergers per year. From this data, it will become possible to perform statistical analyses of the black holes that exist in the universe. For example, one possibility would be to measure the black hole mass spectrum, from which the black hole initial mass function (BHIMF) could be inferred. Both the current and initial black hole mass functions measure the number of black holes as a function of their mass, but the BHIMF is does not include any affects after formation of the black holes, such as accretion or merging. The observational black hole spectrum does include these post-formation effects. 

\begin{figure}
    \centering
    \includegraphics{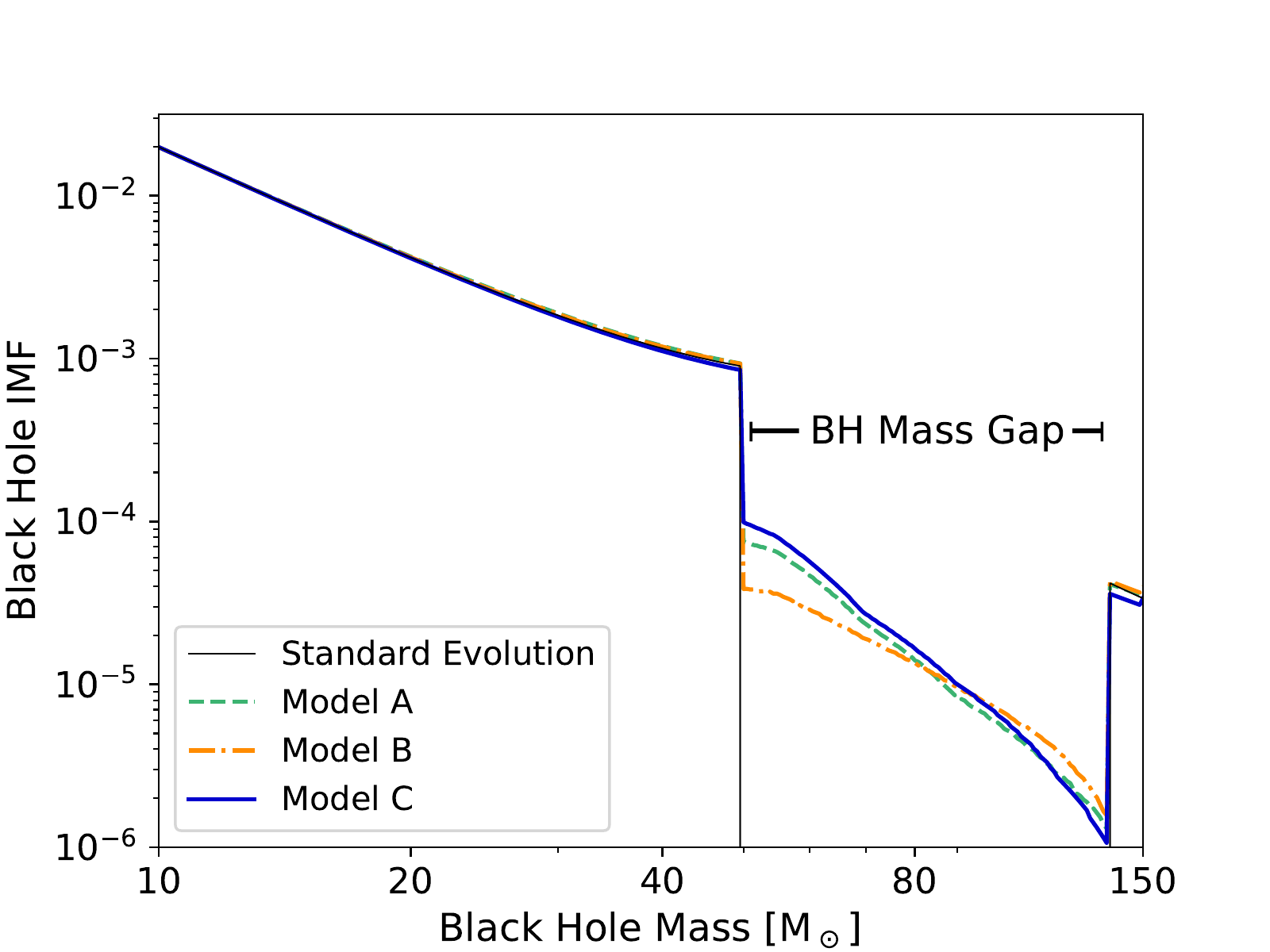}
    \caption{Toy models for the kinds of black hole initial mass function possible by including NNE sources. The thin black curve represents a simplified BHIMF from standard stellar evolution, based on a Salpeter stellar IMF and the mass gap derived from our simulations with 0 NNE. Each of the colored curves represent the BHIMF predicted based on a Salpeter stellar IMF and one of the toy models for the number of stars which contain different amounts of NNE described in the main text. 
    The Salpeter IMF, which provides a basis for all of these BHIMFs, leads to a general decrease in the number of black holes with increasing mass. On top of this general decrease, a small bump in the number of black holes with masses around 40 $M_\odot$ arises because both PPISNe and core collapse supernovae can both produce black holes with masses below the black hole mass gap, leading to a buildup of black holes with those masses. 
    Each model produces different effects in the mass gap, but each case can produce black holes with masses throughout the mass gap, suggesting that it is indeed possible to close the mass gap provided the right conditions.} 
    \label{fig:bhimf}
\end{figure}

In the remainder of this section, we illustrate how our studies of stars containing NNE sources may be used to predict a BHIMF, which could then be compared against future observations. We examine three toy models for the energy injection, chosen for illustrative purposes. The resulting BHIMFs are shown in Figure 5. In each toy model, we approximate the black hole mass ($M_{BH}$) that arises from a star with stellar mass $M_*$ in the following piecewise manner, dependent on the type of evolution we predict it undergoes:
\begin{equation}
M_{BH} = \begin{cases}
    \frac{M_*}{2} & core-collapse\\
    0 & PISN\\
    \frac{M_{*, cc}(E_{NNE})}{2} \left( \frac{M_{*,PISN}(E_{NNE}) - M_*}{M_{*,PISN}(E_{NNE}) - M_{*,cc}(E_{NNE})}\right) & PPISN 
\end{cases}.
\label{eqn:mbh}
\end{equation} In the PPISN case, $M_{*,cc}(E_{NNE})$ and $M_{*,PISN}(E_{NNE})$ correspond to the highest stellar mass of a star that undergoes core collapse and the lowest stellar mass of a star that undergoes pair instability, respectively, for a specified amount of NNE added. In this way, we treat the black holes that arise from PPISN as linearly falling from $M_{*,cc}/2$ to 0 as the mass of the star increases from $M_{*,cc}$ to $M_{*,PISN}$. The factor of 1/2 that appears is a heuristic for the mass lost during the core collapse supernova, used solely for illustrative purposes. A more complete modeling of the supernova is necessary to determine how accurate this heuristic is.

To calculate the number of black holes of a given mass we expect to see, we begin with a Salpeter stellar IMF: $N_*(M) \propto M^{-2.35}$\cite{salpeter1955}. One can see that this slope remains imprinted onto the resulting BHIMF at the end of stellar evolution. To determine the number of black holes of a given mass for a specific amount of added NNE ($N_{BH}(M, E_{NNE}$), we first invert equation~\ref{eqn:mbh} to find a relation $M_*(M_{BH})$. We then plug in this expression to the Salpeter IMF to calculate the number of black holes $N_{BH}(M, E_{NNE}).$ Finally, in order to calculate a total BHIMF, we integrate over the added NNE, with an appropriate weighting $f_{NNE}$:
\begin{equation}
    BHIMF = \int dE_{NNE} f_{NNE}(M, E_{NNE}) N_{BH}(M, E_{NNE}).
\end{equation}
The weighting accounts for what fraction of stars of each mass are injected with a given amount of NNE.

The three toy models we consider differ in the relationship between stellar and black hole mass (equation~\ref{eqn:mbh}) and in the weighting factor $f_{NNE}$. In particular, the masses $M_{*,cc}$ and $M_{*,PISN}$ depend on the boundaries between each of the evolution outcomes. In fact, each of the toy models we consider is based off of one of the contour plots we discuss in section~\ref{results}, as plotted in figures~\ref{fig:summary1}, \ref{fig:summary2}, and \ref{fig:summary3}. Likewise, the weighting factor $f_{NNE}$ also comes from these plots, by dividing the range of added NNE (as described in each plot) into equal-width bins. More specifically, the three models can be summarized:
\begin{itemize}
\item Model A, based on figure~\ref{fig:summary1}: Binned energy rate: We divide the range of NNE rates ($\epsilon$) from 0 to $3.5 \times 10^4 \,  \mathrm{erg\,g^{-1}\,s^{-1}}$ into 200 evenly spaced bins. We then allow 0.1\% of all stars of a given mass to fall into each bin, so that a total of 20\% of stars have some amount of NNE. 
The remaining 80\% of stars are added to the lowest bin. That is, we assume that most stars will have essentially 0 NNE, but if a star has NNE, it is equally likely to have any energy production rate between the the two extremes.
\item Model B, based on figure~\ref{fig:summary2}: Binned integrated energy: We divide the range of integrated energy rates from 0 to $2\times10^{40} \mathrm{erg\, s^{-1}}$ into 200 evenly spaced bins. We allow 0.1\% of stars to fall into each of these bins.  Here, most stars have 0 NNE, but if they have NNE they are equally likely to have any luminosity from NNE.
\item Model C, based on figure~\ref{fig:summary3}: Binned fraction: We divide the fraction of main sequence energy provided by NNE (from 0 to 70\%) into 200 evenly spaced bins. We allow 0.1\% of stars to fall into each of these bins, so that while most stars have 0 NNE, of those that do, the NNE is equally likely to contribute any fraction of the luminosity.
\end{itemize}

Using this approach, we can calculate three examples of the BHIMF we might encounter if stellar evolution includes NNE. These three examples are shown in figure~\ref{fig:bhimf}, along with the BHIMF we would expect from standard stellar evolution (thin black line). Notably, the standard stellar evolution exhibits a mass gap in the range of approximately $45 - 105 M_\odot$, but all three of the models we consider fill this mass gap entirely. That is, in each case, we would expect black holes of all masses to be detectable in gravitational wave observatories.

While these three BHIMFs show significant similarities, there are also differences between these toy models. With sufficient observations, differences like these may be observable, potentially allowing gravitational wave observatories to distinguish between different sources of NNE, or to distinguish NNE from other potential explanations for changes to the BHIMF. However, population modeling would be necessary to determine the weighting factor $f_{NNE}$ for a given physical model of non-nuclear energy, and supernova modeling would be necessary to relate the initial stellar mass to the final black hole mass. Both of these modeling aspects are beyond the scope of this work.

\section{Conclusions}
\label{conclusion}

Through this work, we have shown that introducing non-nuclear energy (NNE) to provide a fraction of the support for the star, can potentially completely fill the black hole mass gap expected due to pair instability supernovae. Adding this NNE source to stars of different masses, ranging from 70 M$_\odot$ to 300 M$_\odot$, can lead to supernova progenitors which could produce black holes with masses at any mass in the mass gap. We were also able to show that for a given model for the fraction of stars affected by this new source of energy, it is possible to predict the black hole initial mass function. From this, it would then be possible to predict the mass spectrum of black holes detectable by current and future gravitational wave observatories, but we leave this calculation to future work.

While most of our results do not assume any particular source of the NNE we add, we contend that dark matter annihilation could provide such a source. In that case, future observations of black holes with masses in the mass gap could offer a powerful new way to probe dark matter. 

\begin{acknowledgements}
K.F. is Jeff \& Gail Kodosky Endowed Chair in Physics at the University of Texas at Austin, and K.F. and J.Z. are grateful for support via this Chair. K.F. and J.Z. acknowledge support by the U.S. Department of Energy, Office of Science, Office of
High Energy Physics program under Award Number DE-SC-0022021 as well as support from the Swedish Research Council (Contract No. 638-2013-8993).
\end{acknowledgements}

\bibliography{Partial_Dark_Star}

\end{document}